\documentclass[fleqn,usenatbib]{mnras}

\usepackage{newtxtext,newtxmath}

\usepackage[T1]{fontenc}

\DeclareRobustCommand{\VAN}[3]{#2}
\let\VANthebibliography\thebibliography
\def\thebibliography{\DeclareRobustCommand{\VAN}[3]{##3}\VANthebibliography}

\usepackage{graphicx}	
\usepackage{amsmath}	
\usepackage{geometry}
\usepackage{array}
\usepackage{booktabs}
\usepackage{subcaption}
\usepackage{dcolumn}
\usepackage{hyperref}
\usepackage{fix-cm}
\usepackage{hyperref}
\usepackage{bm}  

\title[Brightness and colour  variability of NLSy1s]{Brightness and colour variability in NLSy1s} 
\author[Madhu Sudan et al.]{
Madhu Sudan$^{1}$\thanks{E-mail: madhusudan.phdastro@gmail.com},
Hum Chand$^{1}$,
Paul J. Wiita$^{2}$,
Ritish Kumar$^{1,3}$
\\
$^{1}$Department of Physics and Astronomical Science, Central University of Himachal Pradesh (CUHP), Dharamshala 176215, Himachal Pradesh, India\\
$^{2}$ Department of Physics, The College of New Jersey, 2000 Pennington Rd., Ewing, NJ 08628-0718, USA\\
$^{3}$ Inter-University Centre for Astronomy \& Astrophysics, Post Bag 04, Pune, India 411007
}

\date{Accepted XXX. Received YYY; in original form ZZZ}

\pubyear{2024}

\begin{document}
\label{firstpage}
\pagerange{\pageref{firstpage}--\pageref{lastpage}}
\maketitle

\begin{abstract}
The Zwicky Transient Facility (ZTF), with its extensive optical monitoring capabilities, has provided an unprecedented opportunity to study the long-term variability of active galactic nuclei (AGNs). In this work, we present a comparative analysis of optical colour and brightness variability for two $\mathrm{log(L_{Bol})}-z$ matched samples, consisting of 2095 Narrow-line Seyfert 1 (NLSy1) galaxies and a control sample consisting of 2380 Broad-line Seyfert 1 (BLSy1) galaxies. Using over six years of r-band and g-band light curves from the ZTF Data Release 22 (DR22), we characterize flux variability, fractional flux variability, and amplitude of temporal variability for each source in the samples. Our results indicate that BLSy1 galaxies exhibit significantly stronger variability compared to NLSy1s. To probe colour variability, we utilize quasi-simultaneous light curves, with half-hour epoch differences between $\mathrm{g}$- and $\mathrm{r}$-band measurements. We evaluated the colour index using both flux-flux space analysis and linear regression in magnitude-magnitude space. We find that large majorities of these sources --- 74\% of NLSy1 and 79\% of BLSy1 --- exhibit a clear ``bluer-when-brighter'' (BWB) trend, although part of this effect may arise from contamination by the non-varying, predominantly redder flux of the host galaxy. Furthermore, rest-frame structure function analysis reveals that BLSy1 galaxies are $1.44 \pm 0.06$ times more variable than NLSy1s. 
These results can provide valuable insights into the variability properties of AGN subclasses and their underlying physical drivers. 
\end{abstract}

\begin{keywords}
accretion, accretion disks - general: galaxies - galaxies: active - galaxies: Seyfert 1:
techniques: photometric
\end{keywords}


\section{INTRODUCTION}
Active Galactic Nuclei (AGNs) rank among the universe's most luminous phenomena, emitting bolometric luminosities as high as $10^{48} \, \text{erg} \, \text{s}^{-1}$, powered by the accretion of material onto supermassive black holes (SMBHs) \citep[e.g.,][]{1969Natur.223..690L, 1984ARA&A..22..471R, 2002ApJ...579..530W}, and are often modeled using the optically thick, geometrically thin accretion disc framework of \citet{1973A&A....24..337S}. The intensity variation on diverse timescales across the electromagnetic spectrum is a key feature of AGNs. This has been used to probe the emission mechanism around the central engine on scales that are otherwise not resolvable with current or even future technology \citep{1995ARA&A..33..163W, 1997ARA&A..35..445U}.

The classical accretion disc model predicts that fluctuations in the mass accretion rate evolve over viscous timescales, typically hundreds to thousands of years for SMBHs \citep{2018MNRAS.480.3898N}. Yet, observations reveal AGN variability on far shorter scales of months to years in optical and UV bands \citep{2010ApJ...721.1014M}, which are incompatible with the standard mechanism of propagating mass accretion rate fluctuations through a viscous disc, as predicted by classical accretion disc theory \citep{2018NatAs...2..102L}. Modeling of the observed long-term optical variability has been used to either support or refute various possible models, such as disc instabilities \citep{1998ApJ...504..671K}, supernova events \citep{1997MNRAS.286..271A}, and microlensing \citep{2000A&AS..143..465H}. In view of these various possible explanations and/or perhaps due to dependence of variability patterns on various physical parameters such as luminosity, redshift, and black hole mass, the driving physics of variability remains unclear in various subclasses of AGNs.

Among the various subclasses of AGNs, Narrow-line Seyfert 1 (NLSy1) galaxies are relatively rare and are typically characterized by narrow H$\beta$ emission lines (FWHM $< 2000 \, \text{km} \, \text{s}^{-1}$) weak [O~III] (with [O~III]/H$\beta <3$) and strong Fe\,\textsc{II} lines \citep{1985ApJ...297..166O, 1989ApJ...342..224G, 2001A&A...372..730V}.  Although the original definition included strong Fe\,\textsc{II} emission, this criterion was not applied in the recent catalog by \citet{2024MNRAS.527.7055P}, which we used to construct our sample. The threshold of FWHM(H$\beta$) = 2000 km s$^{-1}$ serves as the conventional criterion separating NLSy1 from BLSy1 galaxies, although this division is not based on a fundamental physical distinction.
Compared to BLSy1 galaxies, NLSy1s tend to exhibit steeper X-ray spectra and heightened soft X-ray variability \citep{1999ApJS..125..297L}, typically associated with smaller black hole masses ($10^6 - 10^8 \, M_{\odot}$) and higher Eddington ratios \citep[e.g.][]{2006ApJS..166..128Z}. In terms of their physical properties, NLSy1s are thought to occupy in a distinct region of the AGN parameter space-characterized by low black hole mass, high accretion rate, and often radio-quiet nature, which are in contrast with classical blazars, which typically have high black hole masses, lower Eddington ratios, and strong relativistic jets \citep{2002ApJ...565...78B}.

Like other subclasses of AGNs, NLSy1s also show variability on diverse timescales, but unlike blazars and QSOs, for which variability has been extensively used to probe and constrain their emission mechanisms, large sample-based studies for NLSy1 galaxies are still limited. A relatively recent and extensive study was carried out by \citet{2017ApJ...842...96R}, where they used Catalina Real-Time Transient Survey (CRTS) data to compare the optical variability of NLSy1 and BLSy1 over 5–9 years by modeling light curves using a damped random walk (DRW) to estimate their amplitude of variability. As a class, they found NLSy1 galaxies show a lower amplitude of variability than the BLSy1s. It may be noted that CRTS, with its typical sensitivity corresponding to a limiting magnitude in V $\sim$ 20 in a single exposure and a cadence of about 10 days, makes these light curves noisy and poorly sampled in comparison to those of the ongoing Zwicky Transient Facility (ZTF) \citep{2019PASP..131a8002B}. The ZTF data release DR22\footnote{\url{https://irsa.ipac.caltech.edu/data/ZTF/docs/releases/dr22/ztf_release_notes_dr22.pdf}} covers a baseline of 76 months, with a median cadence of $\sim$3 days in the $g$, $r$, and $i$ bands, resulting in an extensive dataset that is ideal for statistical investigations of AGN variability.

In addition to brightness variability, ZTF is also suitable for studying colour variation over long timescales. For instance, ZTF $g-$ and $r-$band colour variations have been recently studied by \citet[][]{2022MNRAS.510.1791N}, using a large sample of BL Lacs and Flat Spectrum Radio quasars (FSRQs). Their study shows that BL Lacs predominantly follow the bluer-when-brighter (BWB) trend, while the majority of FSRQs exhibit the redder-when-brighter (RWB) trend. Similar colour variation studies for NLSy1s using large samples have not yet been performed, even using CRTS monitoring. The only colour variation study based on a large sample is in the mid-IR and was carried out by \citet{2019MNRAS.483.2362R}. They used the WISE\footnote{\url{https://wise2.ipac.caltech.edu/docs/release/allwise/}} W1–W2 colour and showed that most sources follow the BWB trend.
A similar colour variation study of NLSy1s in the optical, based on the available ZTF $g-r$ colour data, is clearly worthwhile. Similarly, previous large sample size colour variability studies for QSOs, such as the one by \citet{2012ApJ...744..147S} using the SDSS Stripe 82 data, are also limited in comparison to the extensive long-term, and high cadence, $g-r$ colour monitoring available now with ZTF.

Therefore, the recent ZTF-DR22 survey data \citep{2019PASP..131a8002B}, with its extensive optical monitoring capabilities, provides an unprecedented opportunity to study the long-term variability of AGNs. ZTF offers intensive multi-year timescale data, making it an excellent survey for quantifying both optical brightness and colour variability. Moreover, it allows for the construction of controlled samples for various AGN subclasses. In this work, we create matched samples for NLSy1 and BLSy1 galaxies, 
utilizing ZTF's long-term monitoring data spanning more than 6 years. The systematic investigation of the brightness and colour variability properties of these samples using the extensive ZTF datasets is the goal of this work.

The paper is organized as follows. Section~\ref{Data_Preparation} describes the preparation of the samples and the photometric data used for the analysis. Section~\ref{ANALYSIS_AND_RESULTS} presents the analysis methods and results. In Section~\ref{DISCUSSION}, we discuss our findings, followed by a summary and conclusions in Section~\ref{Conclusions}.

\section{DATA AND SAMPLE SELECTION} \label{Data_Preparation}
\begin{figure}
    \centering
    \includegraphics[width=0.48\textwidth, height=0.38\textwidth]{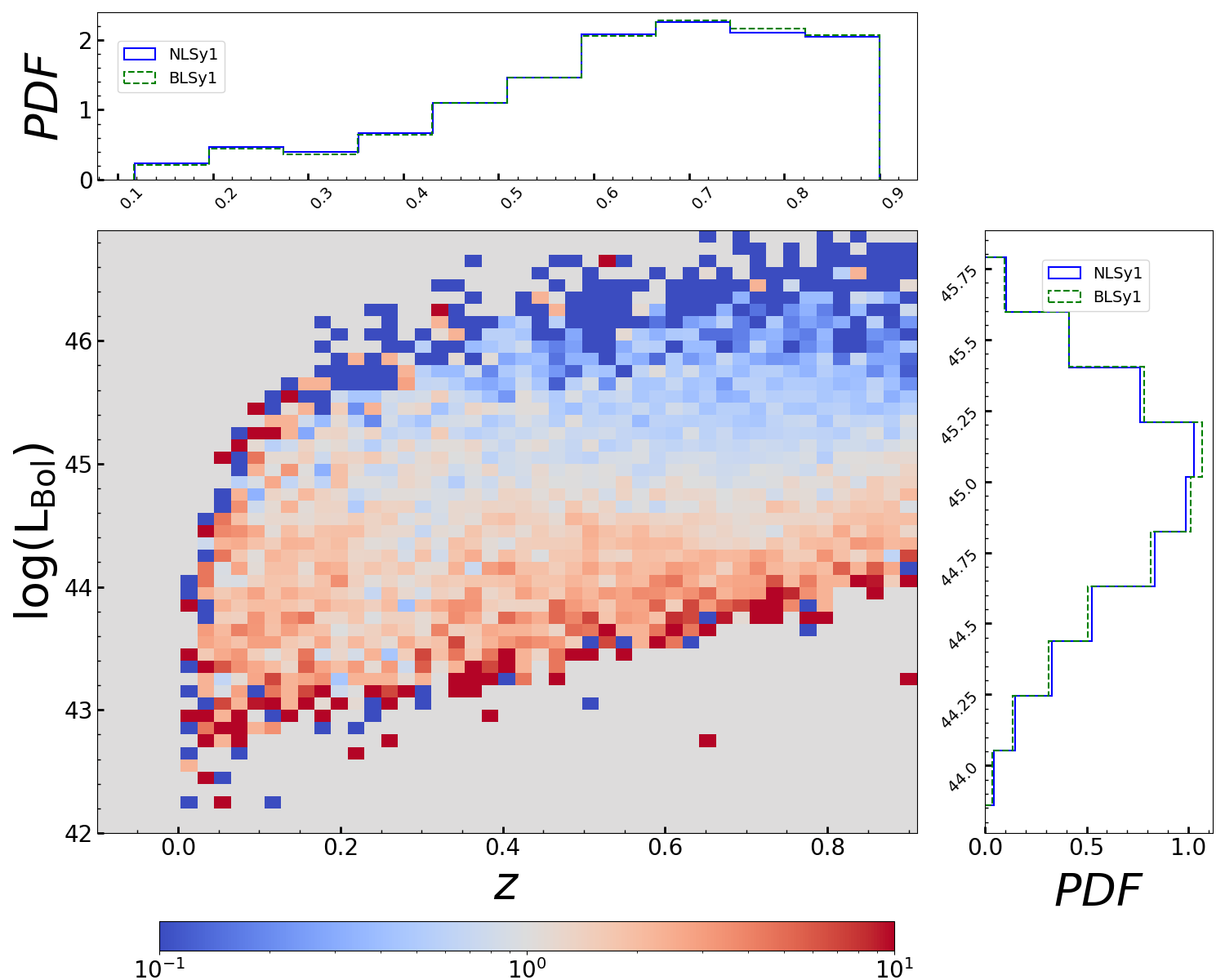}
    \caption{
        Density ratio of NLSy1 and BLSy1 sources matched in the $z$--$\log L_{\mathrm{Bol}}$ plane. 
        The colour code in central panel shows the density ratio of NLSy1 to BLSy1 sources based on normalized 2D histograms. 
        Red and blue regions indicate dominance by NLSy1 and BLSy1 populations, respectively. 
        The top and right panels show the normalised redshift and luminosity histograms for matched NLSy1 and BLSy1 sources. 
        Matching was performed by selecting sources within a 68\% density contour of the NLSy1 distribution and applying tolerances of $\Delta z \leq 0.001$, and $\Delta \log L_{\mathrm{Bol}} \leq 0.001$. The colourbar below the main panel represents the NLSy1/BLSy1 density ratio ranging from 0.1 to 10.}
    \label{fig:nlsy1_blsy1_density_ratio}
\end{figure}

For this study, we began with a sample comprised of 22,656 NLSy1 and 52,723 BLSy1 galaxies from \citet{2024MNRAS.527.7055P}. Their catalog is compiled from the SDSS DR-17 \citep{2017AJ....154...28B}, by decomposing the SDSS  spectra to classify them in different categories viz., NLSy1 with FWHM < 2000 km s$^{-1}$ and BLSy1 with FWHM >2000 km s$^{-1}$. 
For a robust comparative analysis, we selected only those NLSy1 and BLSy1 galaxies that satisfy both of the following criteria: the absolute difference in their redshifts ($\Delta z$) is $\leq 0.001$, and the absolute difference in their bolometric luminosities [$\Delta\log(L_{\mathrm{bol}}/\mathrm{erg~s}^{-1})$] also is $< 0.001$. This ensures that the matched sources lie within a very tight tolerance in both $z$ and $\log(L_{\mathrm{bol}})$, i.e., they are closely paired in the $z$–$\log(L_{\mathrm{bol}})$ plane. Additionally, we restrict the selection to sources lying within the 68\% density contour of the NLSy1 distribution in that space, ensuring a statistically significant sample from the most populated region, as shown in 
Figure~\ref{fig:nlsy1_blsy1_density_ratio}. This process results in initial $\mathrm{log(L_{\mathrm{Bol})}}-z$ matched samples of 2,846 NLSy1 and 3,178 BLSy1 sources.
The distribution of the 68\% density contour of NLSy1s and the corresponding control sample matched in the $L_{\mathrm{Bol}}-z$ plane are illustrated in Figure~\ref{fig:nlsy1_blsy1_density_ratio}.  

Following the construction of this main set of samples, we {searched} for the availability of light curves for our selected sources in the Zwicky Transient Facility (ZTF) DR22 public data release\footnote{\url{https://www.ztf.caltech.edu/}}\citep[see also, e.g.,][]{2019PASP..131a8003M}, which spans approximately six years ($\sim\!76$ months). 

The ZTF assigns separate observation IDs to sources detected under different conditions, as it processes light curves obtained in different fields, filters, and CCD quadrants separately. However, combining light curves from multiple fields and CCD quadrants (even within the same filter) can introduce spurious variability, since ZTF calibrates quadrants independently \citep[see, e.g.,][]{2021AJ....161..267V}. 
To mitigate this issue, we selected the light curve with the highest number of data points associated with each observation ID. Furthermore, we required the \texttt{catflag} parameter to be zero, which indicates that the data were obtained under good observing conditions and are free from known quality issues.

We then impose the criteria that the $\mathrm{r}$-band and $\mathrm{g}$-band light curves in the ZTF DR22 survey have at least 100 data points.
This resulted in a final sample of 2095 NLSy1 and 2380 BLSy1 sources.
The corresponding data are listed in Table~\ref{tab:spectral_data_nlsy1} for NLSy1 and 
Table~\ref{tab:spectral_data_BLSy1} for BLSy1.
The median numbers of data points for NLSy1 and BLSy1 are presented in Table~\ref{tab:analysis_data}, while the median light curve durations are 2217 and 2220 days, respectively.

However, we acknowledge that the samples presented in Tables~\ref{tab:spectral_data_nlsy1}--\ref{tab:spectral_data_BLSy1} may contain spectra exhibiting hybrid features involving combinations of typical Seyfert and starburst galaxy characteristics. To assess the extent of such potential misclassifications, we performed a BPT \citep{1981PASP...93....5B} diagram analysis using narrow emission-line ratios to distinguish between star-forming, composite, and AGN-dominated ionization sources (e.g., see \citealp{2001ApJ...556..121K}; \citealp{2003MNRAS.346.1055K}). For the NLSy1 sample, out of our 2,095 total NLSy1 sources, 405 had measured narrow-line flux values adequate to allow a BPT classification. Among them, 24 were classified as star-forming (5.9\%), 72 as composite (17.8\%), and 309 (76.3\%) as AGN-dominated. Similarly, of the 2,380 BLSy1 sources, 407 had good narrow-line flux measurements; among them, 13 are classified as star-forming (3.2\%), 37 as composite (9.1\%), and 357 (87.7\%) as AGN. This provides us with confidence that contamination from starburst regions is limited and should not have a significant impact on the statistical robustness of our analysis.

\begin{table*}
    \centering
    \caption{Description of the properties in the NLSy1 catalog used in this study. The columns are defined as follows:  
    (1) \texttt{SDSS Name}: SDSS DR17 designation (J2000 coordinates);  
    (2) \textit{$g$-mag}: Median value of $g$-band magnitude corresponding to the source;  
    (3) \textit{$r$-mag}: Median value of $r$-band magnitude corresponding to the source;  
    (4) $N$: Number of data points in $g$- and $r$-band light curves;  
    (5) $z$: Redshift;  
    (6) $\log(H_{\beta,\mathrm{Flux}})$ (erg\,cm$^{-2}$\,s$^{-1}$): Flux of $H_{\beta}$ narrow emission line in logarithmic scale;  
    (7) $\log([\mathrm{O\,III}]_{5007,\mathrm{Flux}})$ (erg\,cm$^{-2}$\,s$^{-1}$): Flux of [O\,III] at 5007\,\AA\ in logarithmic scale;  
    (8) FWHM of broad $H_{\beta}$ line (km\,s$^{-1}$).}
    \label{tab:spectral_data_nlsy1}
    \resizebox{\textwidth}{!}{%
    \begin{tabular}{cccccccc}
      \toprule
        SDSS Name & \textit{$g$-mag} & \textit{$r$-mag} & $N$ & $z$ & $\log(H_{\beta,\mathrm{Flux}})$ (erg\,cm$^{-2}$\,s$^{-1}$) & $\log([\mathrm{O\,III}]_{5007,\mathrm{Flux}})$ (erg\,cm$^{-2}$\,s$^{-1}$) & FWHM H$_{\beta}$ (km\,s$^{-1}$) \\ 
        (1) & (2) & (3) & (4) & (5) & (6) & (7) & (8) \\ 
        \midrule
        
        J021728.28+051432.8 & 20.6 & 20.3 & 346 & 0.5943 & $-16.7977 \pm 0.3634$ & $-16.1099 \pm 0.1029$ & $1619.294 \pm 83.977$ \\
        J075440.10+254449.3 & 19.6 & 19.5 & 956 & 0.6137 & $-16.6436 \pm 0.3439$ & $-15.9161 \pm 0.1328$ & $1117.399 \pm 114.498$ \\
        $-$ & $-$ & $-$ & $-$ & $-$ & $-$ & $-$ & $-$\\ 
        \bottomrule
    \end{tabular}}
    \vspace{0.5cm}
    \textbf{Note:} The full catalog is available in the online version of this paper. This portion of the table is shown for guidance regarding its format and content.
\end{table*}

\begin{table*}
    \centering
    \caption{Same as Table~\ref{tab:spectral_data_nlsy1}, but for the BLSy1 control sample. This sample is matched to the NLSy1 sample in the $\log(L_{\mathrm{Bol}})$--$z$ plane within a tolerance of 0.001.}
    \label{tab:spectral_data_BLSy1}
    \resizebox{\textwidth}{!}{%
    \begin{tabular}{cccccccc}
        \toprule
        SDSS Name & \textit{$g$-mag} & \textit{$r$-mag} & $N$ & $z$ & $\log(H_{\beta,\mathrm{Flux}})$ (erg\,cm$^{-2}$\,s$^{-1}$) & $\log([\mathrm{O\,III}]_{5007,\mathrm{Flux}})$ (erg\,cm$^{-2}$\,s$^{-1}$) & FWHM H$_{\beta}$ (km\,s$^{-1}$) \\ 
        (1) & (2) & (3) & (4) & (5) & (6) & (7) & (8) \\ 
        \midrule
        
        J000003.24+113119.1 & 20.6 & 20.4 & 327 & 0.4819 & $-15.8789 \pm 0.0344$ & $-15.1734 \pm 0.0159$ & $4556.422 \pm 273.200$ \\ 
        J000028.78+301616.2 & 19.9 & 19.5 & 753 & 0.7762 & $-16.6696 \pm 0.2329$ & $-15.6440 \pm 0.0593$ & $6515.089 \pm 496.878$ \\ 
        $-$ & $-$ & $-$ & $-$ & $-$ & $-$ & $-$ & $-$ \\ 
        \bottomrule
    \end{tabular}}
    \vspace{0.5cm}
    \textbf{Note:} The full catalog is available in the online version of this paper. This portion of the table is shown for guidance regarding its format and content.
\end{table*}

\section{ANALYSIS AND RESULTS} \label{ANALYSIS_AND_RESULTS}
In this study, we investigate the brightness variability of AGNs by analyzing flux, fractional flux, temporal variability amplitude, and colour variability, along with a structure function analysis. We compare the variability properties of NLSy1 (2095 sources) with BLSy1 (2380 sources).

\subsection{Flux and Fractional Flux Variability} \label{flux_frac}
The flux variability observed in AGNs is driven by a variety of physical mechanisms operating on different timescales and spatial scales. For radio-quiet AGNs, such as those used in our sample, small-amplitude variations may arise from accretion disc instabilities \citep[e.g.,][]{1991sepa.conf..557W, 1993ApJ...406..420M, 1993ApJ...411..602C}.
Additional mechanisms include thermal-viscous instabilities, changes in accretion rates, and turbulence potentially driven by magneto-rotational instabilities (MRI) \citep[e.g.,][]{1998ApJ...504..671K, 2009ApJ...698..895K, 2018Galax...6..139K}.
Stochastic models such as the damped random walk (DRW) have also been successfully employed to describe optical variability in quasars and Seyfert galaxies \citep{2009ApJ...698..895K, 2010ApJ...721.1014M}.

Following \citet{2010ApJ...716L..31A}, we computed the intrinsic flux variability ($F$), which quantifies the variance of observed light curves, in both the $\mathrm{g}$- and $\mathrm{r}$-bands, while accounting for measurement uncertainties.

Estimation of $F$ follows the methodology described in \citet{2007AJ....134.2236S}.  One can define the gross variation as
 \begin{equation}
\Sigma  = \sqrt{\frac{1}{N-1} \sum_{i=1}^{N}\left(m_{i} - <{m}>\right)^2}~~,
\end{equation}
where $m_{i}$ is the magnitude at the $i^{th}$ measurement in a light curve of $N$ points and $<m>$ is their error weighted average (weighted by $\epsilon_i^{-2}$, with $\epsilon_i$ the error on the $i^{th}$ measurement). Then a way to express the flux variability's amplitude is as
\begin{equation}
    F= 
\begin{cases}
    {\Sigma ^2-\epsilon^2},& \text{if } \Sigma > \epsilon \\
    0,              & \text{otherwise}.
\end{cases}
\label{flux_variability}
\end{equation}
Here $\epsilon$ is the measurement error contribution to variance, and it is directly computed from the errors of each unique observed magnitude $\epsilon_i$ as,
\begin{equation}
\epsilon^2 = \frac{1}{N} \sum_{i=1}^{N} \epsilon_i^2~.
\end{equation}

\begin{table}
    \centering
    \caption{Comparison of variability properties among NLSy1 and BLSy1. Listed quantities include median values both in $\bm{\mathrm{g}}$- and $\bm{\mathrm{r}}$-band, for the number of data points (<N>) and time duration (<T>) of light curves in days, along with median flux variability (<F>), temporal amplitude variability $<\Psi>$, and fractional variability ($F_{\text{var}}$) with uncertainties. The median difference in fractional variability between the $g$- and $r$-bands is given by $<F_{\text{var,g}} - F_{\text{var,r}}> \equiv <\Delta F_{\text{var,gr}}>$.}
    \label{tab:analysis_data}
    \footnotesize 
    \renewcommand{\arraystretch}{1.2} 
    \setlength{\tabcolsep}{4pt} 
    \begin{tabular}{ccc}
        \toprule
        \textbf{Quantities} & \textbf{NLSy1} & \textbf{BLSy1} \\
                             &  & \text{($\log(L_{\mathrm{Bol}})$--$z$ matched to NLSy1)} \\
        \midrule
         Sample size    & 2095 & 2380 \\
        <$N_{r}$> & 527 & 560 \\
        <$N_{g}$> & 340 & 360 \\
        <$T_{r}$/days> & 2217 & 2219 \\
        <$T_{g}$/days> & 2241 & 2243 \\
        $<F_r>\ (mag^2)$ & 0.0826 & 0.1144 \\
        $<F_g> (mag^2)$ & 0.1033 & 0.1401 \\
        $<F_{\text{var,r}}>$ & $1.921^{+0.227}_{-0.227} \times 10^{-3}$ & $2.615^{+0.161}_{-0.161} \times 10^{-3}$ \\
        $<F_{\text{var,g}}>$ & $1.988^{+0.216}_{-0.216} \times 10^{-3}$ & $2.682^{+0.157}_{-0.157} \times 10^{-3}$ \\
        <$F_{\text{var,g}} -F_{\text{var,r}}$>  & $0.040^{+0.267}_{-0.267} \times 10^{-3}$ & $0.045^{+0.213}_{-0.213} \times 10^{-3}$ \\
        $<\Psi_{r}> (mag)$ & 0.7675 & 0.8325 \\
        $<\Psi_{g}> (mag)$ & 0.8229 & 0.9016 \\
        \bottomrule
    \end{tabular}
\end{table}

\begin{figure}
    \centering
    \includegraphics[width=0.45\textwidth,height=0.3\textwidth]{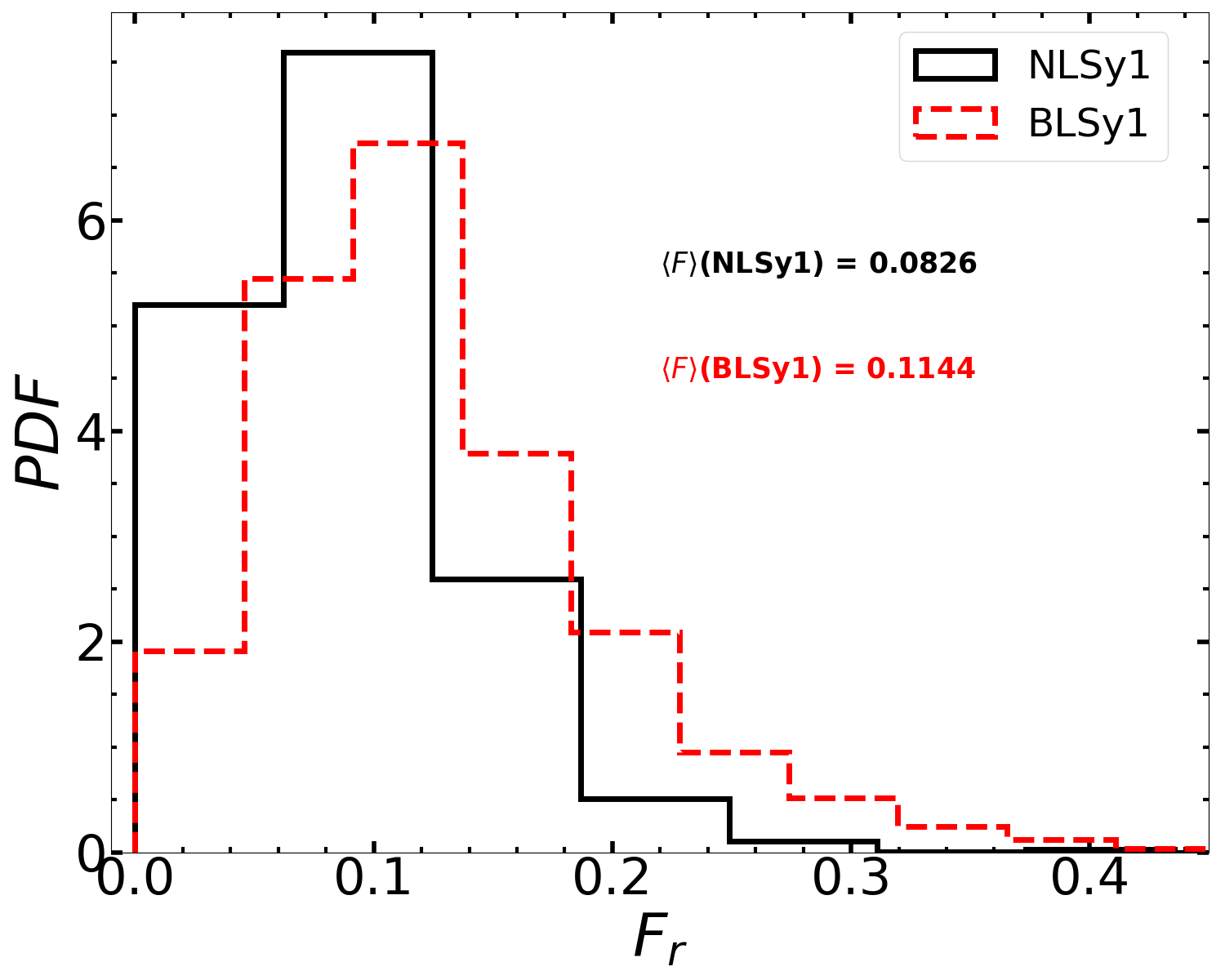}
    \caption{Comparison between NLSy1 and BLSy1 flux variability distributions, demonstrating a significant difference as indicated by the KS2 test, which yields a null p-value of $1.39 \times 10^{-62}$.} 
    \label{fig:Flux_Variability}
\end{figure}

As emphasized in \citet{2007AJ....134.2236S}, the total observed scatter in a light curve includes contributions from both genuine astrophysical variability and photometric uncertainties. By subtracting the mean square measurement error from the observed variance, the quantity $F$ captures the net intrinsic variability. Unlike fractional variability, $F$ retains units of magnitude squared (mag$^2$) and reflects the absolute power of fluctuations, making it suitable for identifying sources with strong intrinsic changes, even at low flux levels. Using the above formulae 
we have calculated the flux variability for all 2095 NLSy1 and 2380 BLSy1.  
Table \ref{tab:analysis_data} gives the median values of flux variability for NLSy1 and BLSy1. 
Hence we see that for these samples matched in luminosity and redshift the BLSy1 sample exhibits more flux variability than does the  NLSy1 sample. 
Plots comparing these flux variabilities in $\mathrm{r}$-band are given in Figure \ref{fig:Flux_Variability}.

Then we calculated the fractional flux variability amplitude ($F_{var}$), to quantify the degree of variability in a light curve, as given by \citet{2003MNRAS.345.1271V}: 
\begin{equation}
     F_{var} = \sqrt{\frac{S^{2} - \epsilon^{2}}{\overline{m^{2}}}}~,
     \label{eq:fvar_equation}
 \end{equation}
where,
\begin{equation}
S  = \sqrt{\frac{1}{N-1} \sum_{i=1}^{N}\left(m_{i} - \overline{m}\right)^2}~~.
\end{equation}

The error in $F_{var}$ is given by
 \begin{equation}
 \epsilon (F_{var}) = \sqrt{\left( \sqrt{\frac{1}{2N} }\frac{\epsilon^{2}}{\overline{m^{2}} F_{var}} \right)^{2} + \left( \sqrt{ \frac{\epsilon^{2}}{N} }\frac{1} {\overline{m\
}}  \right)^{2}} ~~
    \label{eq:fvarerr_equation}
\end{equation}
In the above formulation, the overline notation (e.g., $\overline{m}$) denotes a simple arithmetic mean, i.e., without any weight. 

\begin{figure}
	\includegraphics[width=0.45\textwidth,height=0.3\textwidth]{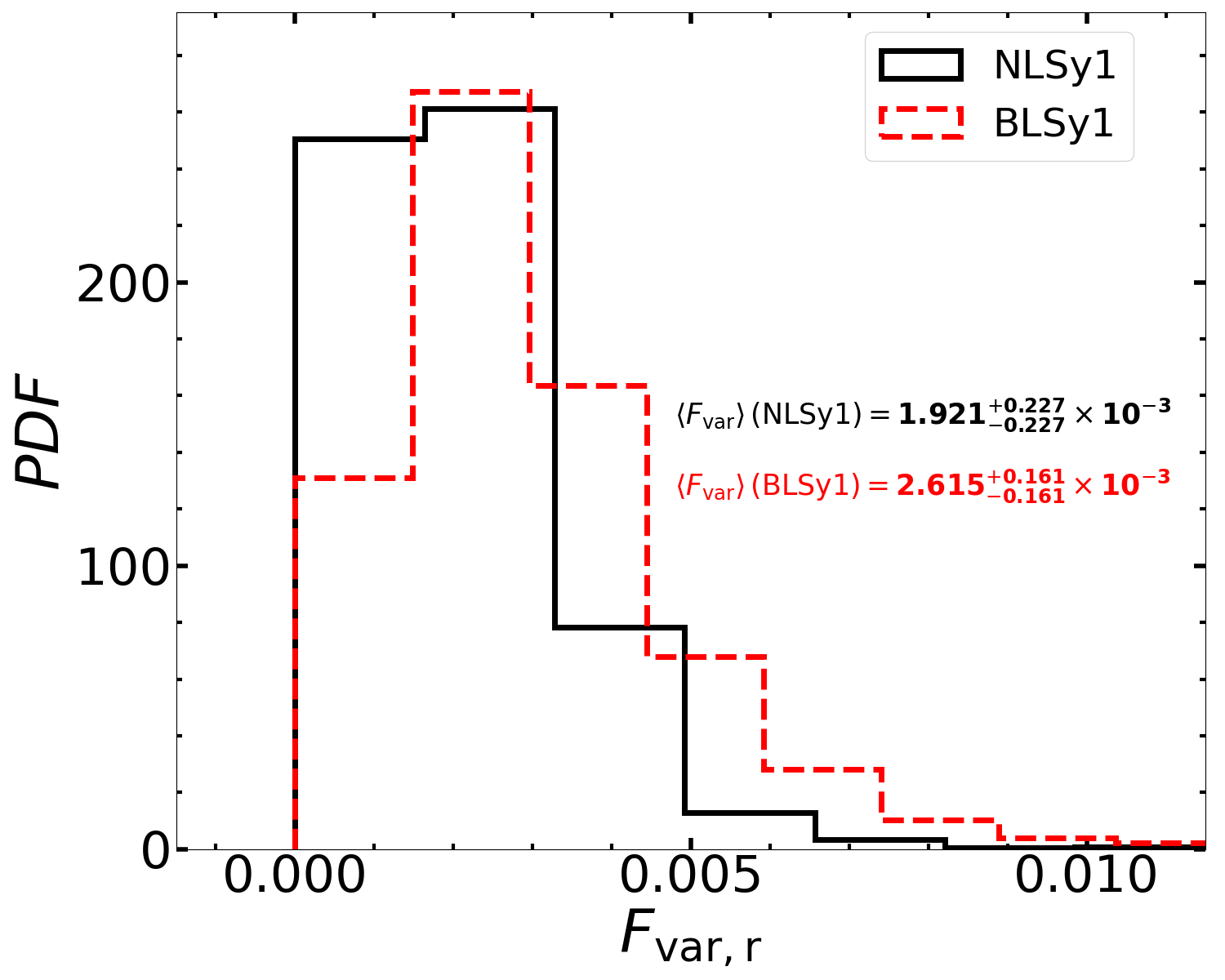}
    \caption{Comparison between NLSy1 and BLSy1 fractional variability distributions, demonstrating a significant difference as indicated by the KS2 test: $p = 3.57 \times 10^{-41}$.}
    \label{fig:Fractional_Variability}
\end{figure}

It can be noted that $F_{var}$ measures the amplitude of flux variations relative to the mean flux, normalized to account for statistical noise. It provides a linear statistic that can express variability in percentage terms, making it easier to interpret. This makes $F_{var}$ particularly useful for comparing variability across sources with different flux levels or for detecting intrinsic variability amplitude changes across different wavebands or epochs. Unlike raw variance, which depends on the absolute flux level, $F_{var}$, is normalized and dimensionless, allowing for a fair comparison. It is important to note that, although $F_{var}$ is straightforward to compute, its statistical properties are influenced by the stochastic nature of AGN variability. 
For the NLSy1 and BLSy1 samples we have calculated the $F_{var}$ and other related quantities in the available $\mathrm{g}$-band and $\mathrm{r}$-band. 
In Figure~\ref{fig:Fractional_Variability}, we present the comparative fractional variability in the $\bm{r}$-band for the NLSy1 and BLSy1 sources, which are found to be $(1.921 \pm 0.227) \times 10^{-3}$ and $(2.615 \pm 0.161) \times 10^{-3}$, respectively.

\begin{figure}
    \centering
    \includegraphics[width=0.45\textwidth,height=0.30\textwidth]{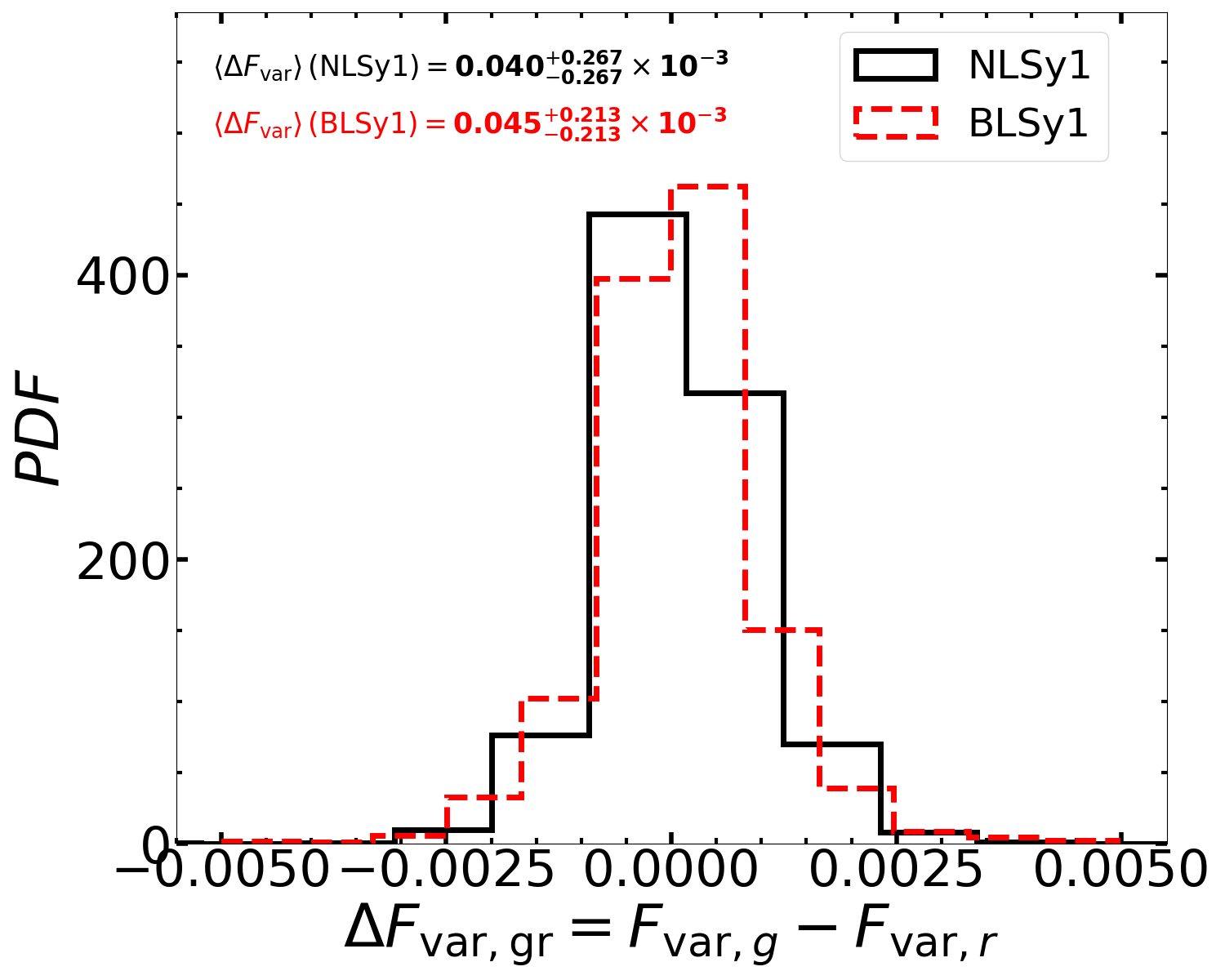}
    \caption{Comparison between the colour differences in fractional variability distributions for the NLSy1 and BLSy1 samples, which show a marginally significant difference, as indicated by the KS2 test with $p = 0.0120$.}
    \label{fig:Delta_Fractional_Variability}
\end{figure}

Next we analyzed the difference in fractional variability between the $\mathrm{g}$- and $\mathrm{r}$-bands and quantified it as $\Delta F_{\text{var, gr}} = F_{\text{var,g}} - F_{\text{var,r}}$. Uncertainty in $\Delta F_{\text{var}}$ has been calculated as $\sqrt{\epsilon(F_{\text{var,g}})^2 + \epsilon (F_{\text{var,r}})^2}$. 
For NLSy1 sources, we found that $F_{\text{var}}$ is equal  in both bands  within 1$\sigma$ uncertainties for 47.92\% (1004/2095),  greater ( at $>1\sigma $) in the $\mathrm{g}$-band for 29.55\% (619/2095), and greater (bf at $>1\sigma $)in the $\mathrm{r}$-band for 22.53\% (472/2095) of them. Similarly, for BLSy1 sources, $F_{\text{var}}$ is equal in 32.10\% (764/2380), greater ( at $>1\sigma $ ) in the $\mathrm{g}$-band for 37.23\% (886/2380), and greater ( at $>1\sigma $) in the $\mathrm{r}$-band for 30.67\% (730/2380) of them. These results suggest that $F_{\text{var}}$ may exhibit a modest difference in frequency-dependence between these groups of sources, with median value of  $\Delta F_{\text{var, gr}}$ consistent with zero,  as shown in Figure~\ref{fig:Delta_Fractional_Variability}. 

\subsection{Amplitude of Temporal Variability} \label{amp_var_sec}
 The amplitude of temporal variability ($\Psi_{r}$) refers to the peak-to-peak amplitude of flux variations observed in the optical light curves. It measures the maximum change in brightness between the highest and lowest flux states. This estimator is particularly effective in identifying sources that undergo strong, isolated outbursts or flaring activity, even if the overall variance of the light curve remains moderate. It provides a straightforward and intuitive quantification of variability that does not depend on the light curve’s statistical distribution or temporal sampling density. As emphasized in \citet{1996A&A...305...42H}, $\Psi$ is especially well-suited for characterizing intra-night or short-term optical variability in AGN and blazars, as it captures the maximum observed excursion corrected for measurement noise. This correction ensures that the measured amplitude reflects intrinsic variability rather than merely observational uncertainty. This amplitudes of temporal variability for all of our sample datasets in the $\mathrm{g}$- and $\mathrm{r}$-band have been calculated as follows, following  \citet{1996A&A...305...42H}:
\begin{equation}
    \Psi_{r,g} = \sqrt{(A_{max}-A_{min})^{2}-2\epsilon^{2}}
    \label{eq:varamp_equation}
\end{equation}
where $\epsilon^{2} = \langle \epsilon_{i}^{2} \rangle$, and $\epsilon_{i}$ indicates the uncertainty in the $i^{th}$ data point. The maximum and minimum amplitudes in the light curve are represented by $A_{max}$ and $A_{min}$, respectively. Unlike $F_{var}$, which provides a normalized measure of flux fluctuations across the entire light curve, $\Psi$ is sensitive to individual extreme events and thus complements other variability diagnostics in revealing transient or episodic phenomena in AGN light curves.

\begin{figure}
    \centering
    \includegraphics[width=0.45\textwidth,height=0.30\textwidth]{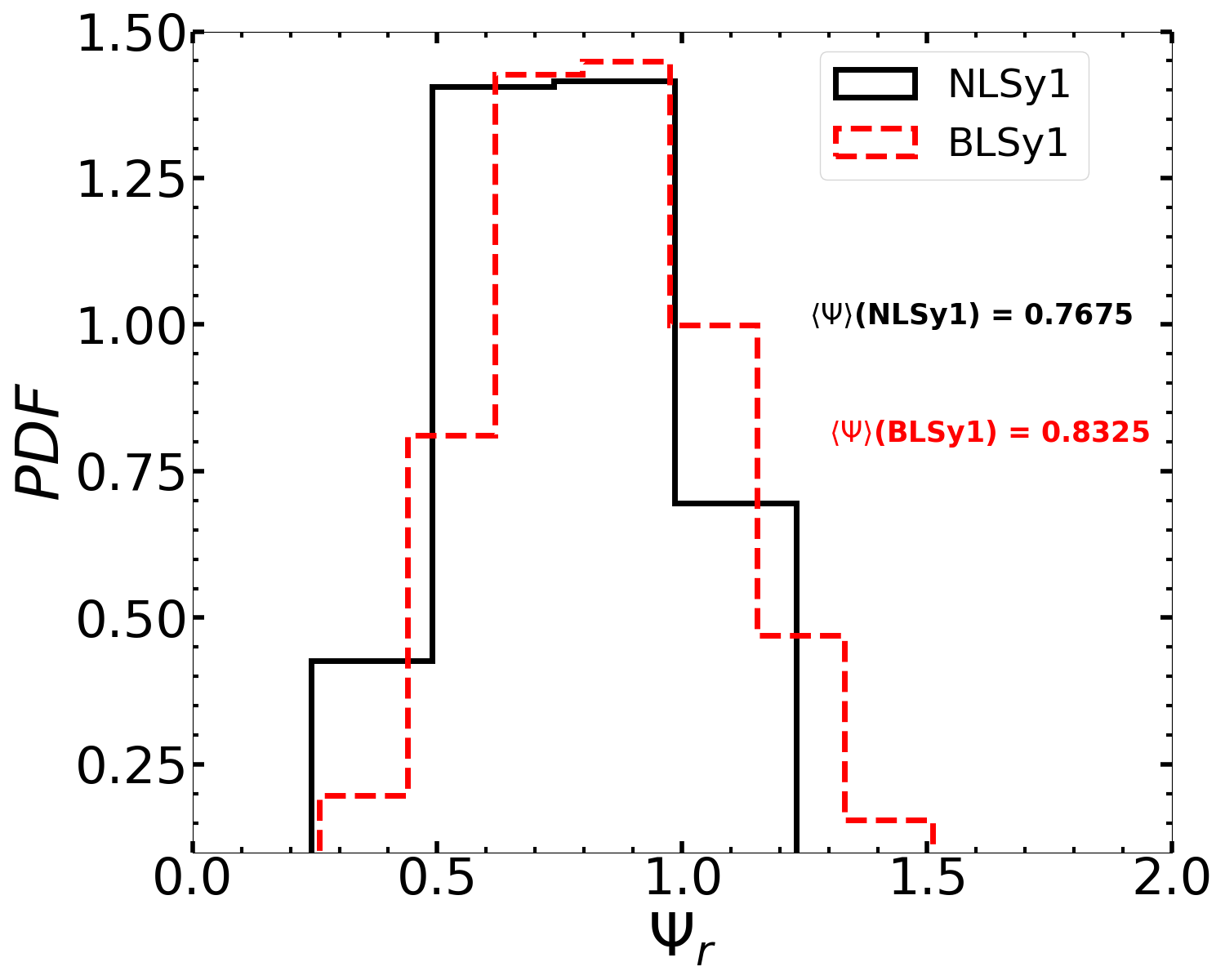}
    \caption{Comparison between the NLSy1 and BLSy1 amplitude of temporal variability distributions in the r-band, demonstrating a significant difference between them as indicated by the KS2 test with $p = 2.15 \times 10^{-14}$.}
    \label{fig:Amplitude_Variability}
\end{figure}

Since both $A_{\mathrm{max}}$ and $A_{\mathrm{min}}$ are sensitive to outliers, we first applied a $3\sigma$ clipping procedure to mitigate their impact. If any outliers remained after this step, we further refined the estimation by sorting the magnitude data in ascending order. We then defined $A_{\mathrm{max}}$ as the median of the five largest magnitude values and $A_{\mathrm{min}}$ as the median of the five smallest magnitude values. Using this method, the median values of $\Psi_{r}$ for the NLSy1 and BLSy1 samples are found to be $0.77$ and $0.83$, respectively (Table~\ref{tab:analysis_data}).
We see that in  both $\mathrm{g}$- and $\mathrm{r}$-bands the BLSy1 sample exhibits a higher median $\Psi$ value than the NLSy1 sample, indicating greater amplitude of temporal variability. 
We find that the amplitude of temporal variability $\Psi$ is higher in the $\mathrm{g}$-band than in the $\mathrm{r}$-band for both Seyfert subclasses (see Table \ref{tab:analysis_data}). These comparative trends in $\mathrm{r}$-band amplitude of temporal variability are  presented in Figure \ref{fig:Amplitude_Variability}.

\subsection{Structure Function Analysis} \label{SF_sec}
\begin{figure*}
    \centering
    \includegraphics[width=0.49\textwidth, height=0.32\textwidth]{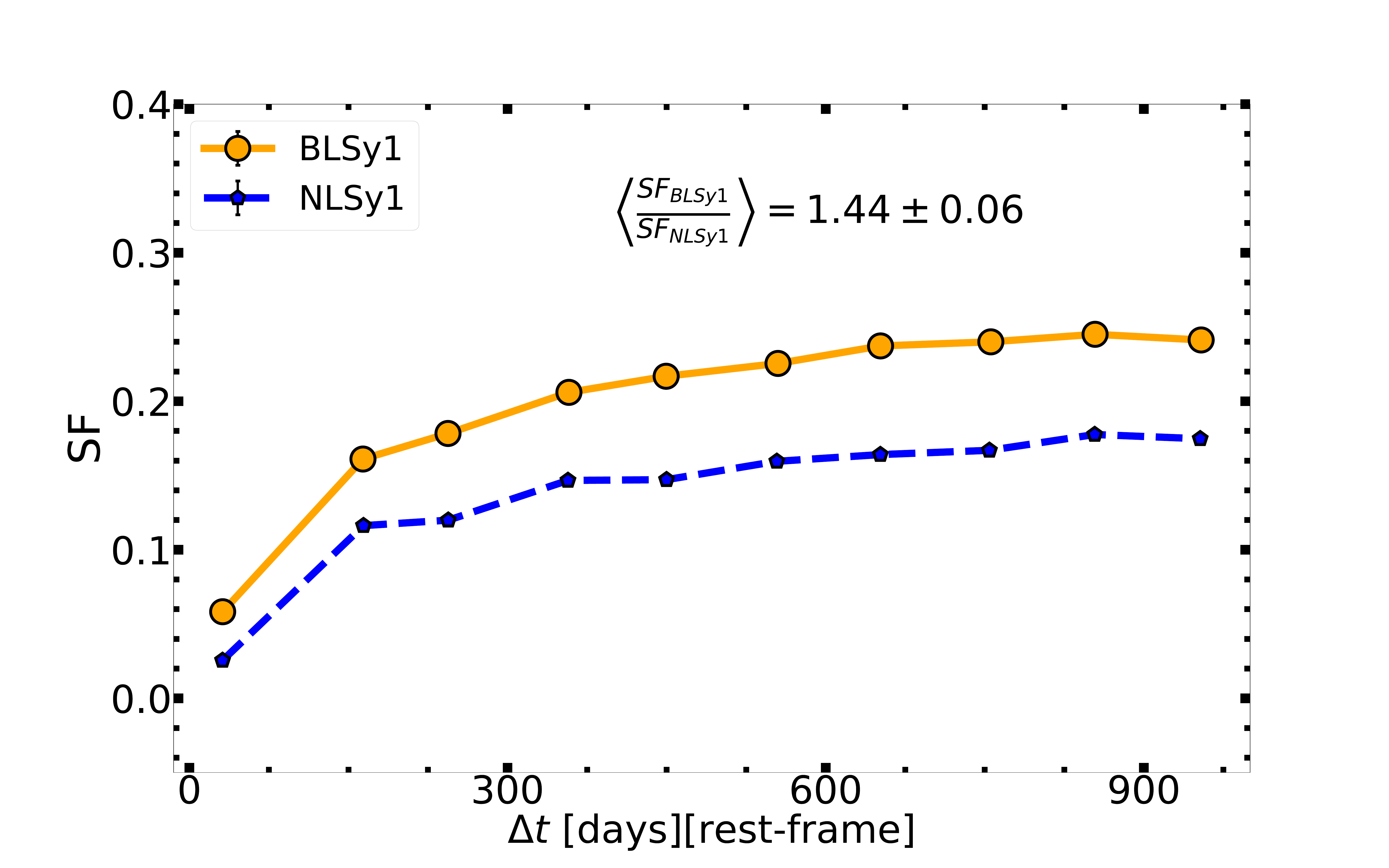}
    \includegraphics[width=0.49\textwidth, height=0.32\textwidth]{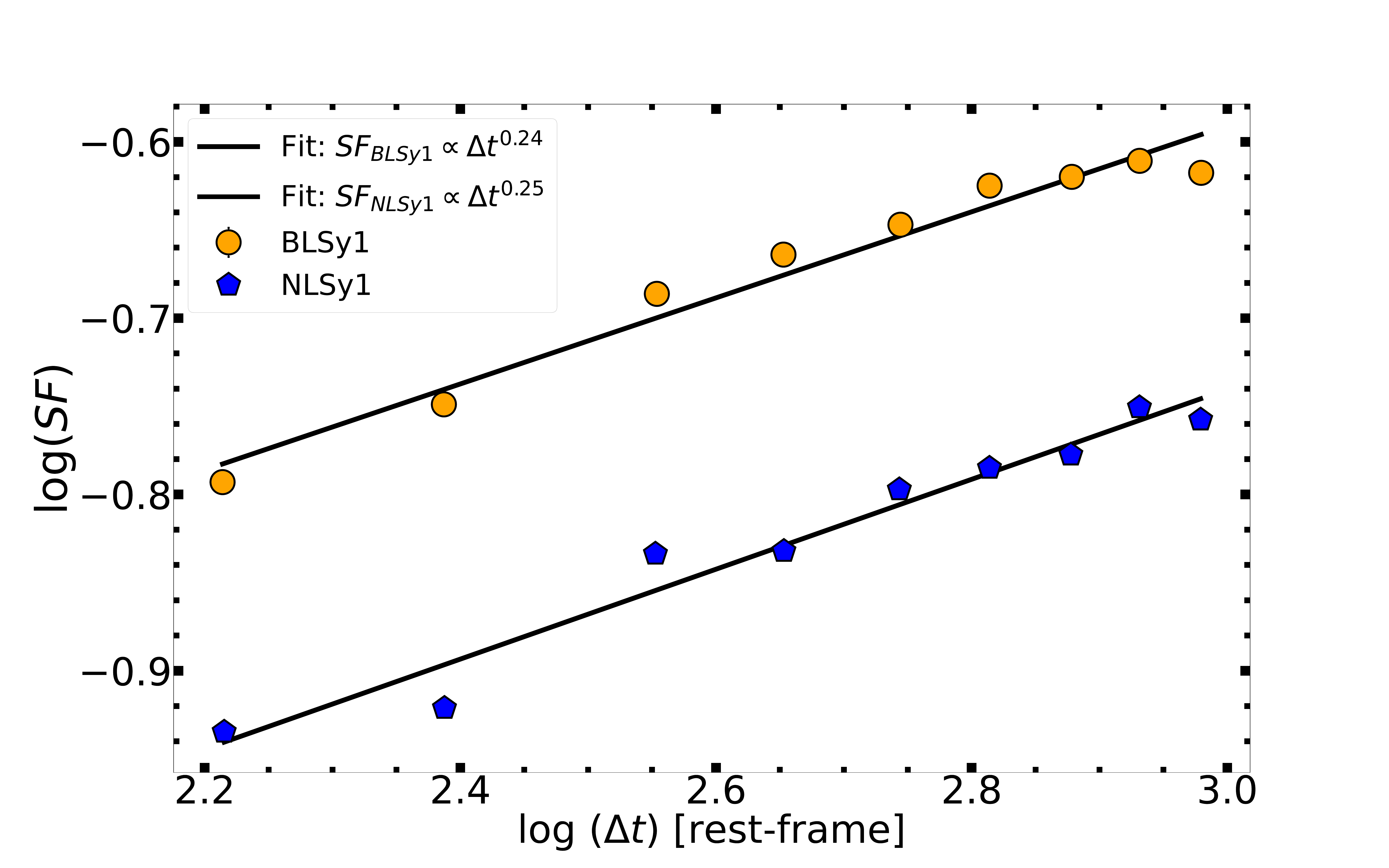}
    \caption{
        Structure function analysis of BLSy1 galaxies (filled orange circles) and NLSy1 galaxies (filled blue pentagons). 
        The plots clearly illustrate that BLSy1 sources exhibit significantly stronger variability than NLSy1 sources, 
        highlighting the distinct variability characteristics of these AGN subclasses. 
        Error bars are smaller than the symbol sizes and are therefore not visible.
    }
    \label{fig:SF_dist_BLSy1}
\end{figure*}

The structure-function (SF) derived for a light curve is often used to infer variability properties such as characteristic time scales and hints of any periodicities. Several definitions of SF have been used in the literature \citep[see e.g.,][and references therein]{2002MNRAS.329...76H,de2005structure, 2011A&A...525A..37M,graham2014novel}. A SF also can be built from the light curves of all the sources in a sample \citep[e.g., see,][]{Ritish2025}. It is a measure of the ensemble root mean square (rms) magnitude difference as a function of the time lag between different epochs, for which the measurements are typically grouped together into bins. Such a statistical approach can be readily applied to large samples of objects, as we consider here, for the characterization of their typical variability properties. The strength of SF analysis lies in the mutual independence of the different bins over which the variability is measured \citep[e.g.,][]{de2005structure}. Here we use the modified SF, as defined in \cite{di1995variability}, 
\begin{equation}
    SF(\Delta t) = \sqrt{\frac{\pi}{2}\langle |m(t+\Delta t)-m(t)|\rangle^2-\langle \sigma_{n}^2 \rangle}~
\end{equation} 
Here $m(t)$ is the time-dependent magnitude, and terms within the $\langle ~~ \rangle $ indicate the time-average taken over the entire light curve. In the first term, the square is performed after the average in order to reduce the influence of possible outliers \citep{1994MNRAS.268..305H}; $\sigma_{n}$ is the contribution due to the noise of the data. The overall noise contribution  can be estimated as the standard deviation in each pair separated by a time lag $\Delta t$. The factor $\frac{\pi}{2}$ is introduced under the assumption that both the intrinsic variability and the noise can be represented via a Gaussian distribution \citep{wilhite2008variability}. The SFs of our NLSy1 sample and the control sample of BLSy1 galaxies are shown in the left panel of Figure \ref{fig:SF_dist_BLSy1}.

This figure illustrates that BLSy1 exhibit greater variability than NLSy1 galaxies across our measured restframe  timescales ranging from 20 to 900 days. Specifically, the median SF value for BLSy1 sources is $1.44 \pm 0.06$ times higher than that of the NLSy1 sample.

In principle, the structure function (SF) serves as the time-domain counterpart of the power spectral density (PSD), offering complementary insights into variability characteristics. Both frameworks exhibit analogous features: a flat region at short timescales dominated by noise, a rising power-law segment indicating correlated variability, and a flattening at longer timescales where the signal becomes uncorrelated—resembling the form of a bending or broken power-law PSD. The slope of the SF quantifies how quickly variability grows with increasing time lag. From the $\log \text{SF}$–$\log \Delta t$ plots shown in the right panel of Figure~\ref{fig:SF_dist_BLSy1}, the best-fit SF slopes are found to be $0.24 \pm 0.02$ for BLSy1 and $0.25 \pm 0.02$ for NLSy1, so they are fully consistent in this characteristic. 

\subsection{Colour variability Analysis} \label{colour_var_sec}
AGN are known to show colour variability when their fluxes change. 
Both  quasi-thermal emission from the accretion disc and non-thermal synchrotron radiation from the relativistic jet affect their optical emission. By examining the samples' colour variations, we can attempt to differentiate between different possible modes responsible for flux variation.
Here, we constrain the colour variation over large periods, ranging from months to years by using quasi-simultaneous observations for these samples, as they have gaps of only about half an hour between $\mathrm{g}$- and $\mathrm{r}$-band measurements. 
\begin{table}
    \centering
    \caption{Results of colour variability based on ZTF $g-r$ band colour. The analysis is conducted in both flux–flux and magnitude–magnitude spaces, with values from the latter presented in parentheses.}
    \label{tab:colr_variability_data}
    \begin{tabular}{ccc}
        \toprule
        Colour trend & NLSy1 & BLSy1 \\
             & & (Matched to \\          
             & & NLSy1) \\     
        \midrule
        Total & 2095 & 2380 \\
        RWB & 161(6) & 140(12) \\
        BWB & 1555(2013) & 1870(2257) \\
        No-trend & 379 (76) & 370(111) \\
        \bottomrule
    \end{tabular}
\end{table}

\begin{figure*}
    \centering
    \includegraphics[width=0.49\textwidth,height=0.32\textwidth]{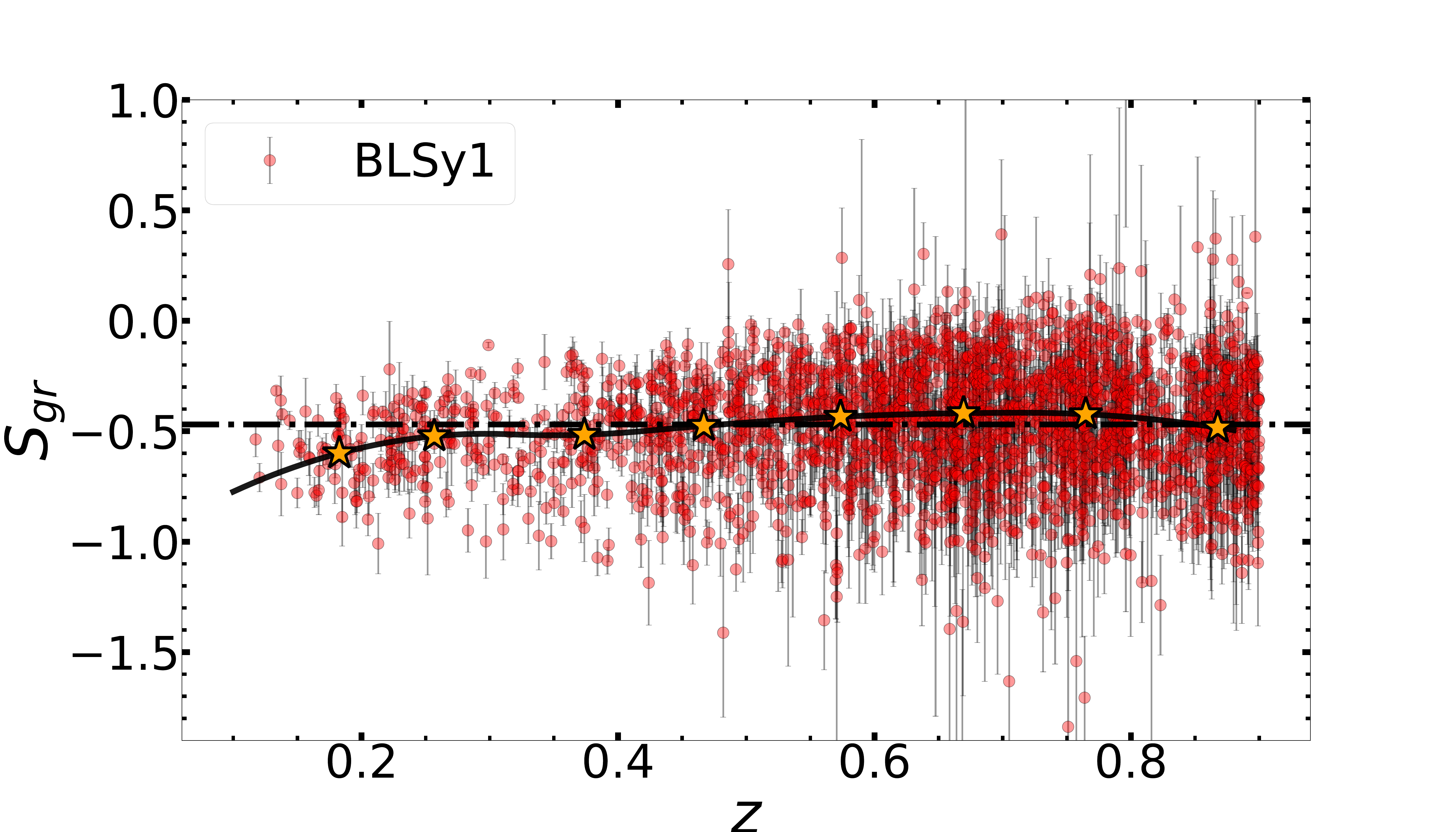}
    \includegraphics[width=0.49\textwidth,height=0.32\textwidth]{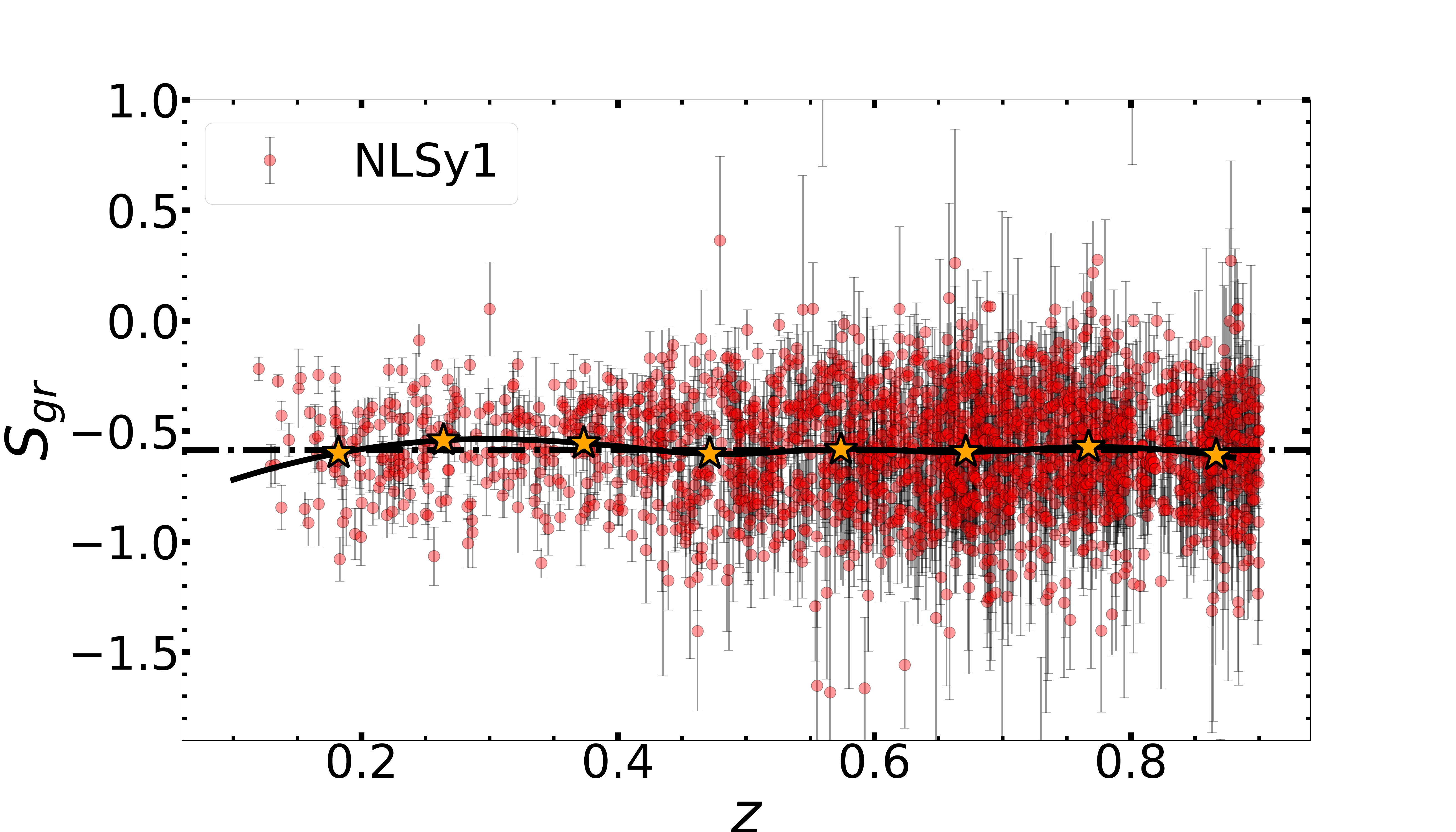}
    \caption{Distribution of the colour variability index, $S_{gr}$, as a function of redshift, $z$, where the solid line represents the smoothed trend line for the sample, and the dashed-dot line marks the mean $S_{gr}$ value for the samples. \textbf{\textit{Left:}} $S_{gr}$ distribution for BLSy1 sources. \textbf{\textit{Right:}} $S_{gr}$ distribution for NLSy1 sources.}
    \label{fig:SF_dist}
\end{figure*}

Some past studies quantified the colour variability by fitting data in colour-magnitude space, for instance, in $g$ versus $(g - r)$ space \citep{8135723, 2004ApJ...601..692V, 2005ApJ...633..638W}; however, this approach  suffers from covariances between the colour and magnitude uncertainties. This drawback has been addressed by \citet{2012ApJ...744..147S} where they fit data in magnitude-magnitude space and then translate it to colour-magnitude space, as follows:

\begin{equation}
r - \langle r \rangle = S'_{gr} (g - \langle g \rangle) + P,
\end{equation}
\begin{equation}
r = S'_{gr} g + Q,
\end{equation}
and
\begin{equation}
g - r = -(S'_{gr} - 1) (g - \langle g \rangle) + R,
\end{equation}
where $\langle r \rangle$  and  $\langle g \rangle$ are average value of the $g$- and $r$-band magnitude in the light curve. Here $S'_{gr}$ denotes the slope of the linear relation, while $P$, $Q$, and $R$ are the corresponding line-fitting constants. Specifically, $Q = \langle r \rangle - S'_{gr} \langle g \rangle + P$ and $R = -P + (\langle g \rangle - \langle r \rangle)$. 
These equations represents three equivalent linear fits, providing the transformation 
between magnitude–magnitude space and colour-magnitude space. Then
\begin{equation}
S_{gr} = (S'_{gr} - 1)
\end{equation}
is the colour index.
To eliminate the redshift dependency from the colour variability behaviour induced by emission lines, we have also corrected the individual values of $S_{gr}$ by following \citet{2012ApJ...744..147S}, in defining
\begin{equation}
     S_{gr, corr}= S_{gr} +  \langle S_{k} \rangle (z) ~  
\end{equation}
with $\langle S_{k} \rangle (z)$  being the redshift dependent  deviation.  These are seen as the differences between the dot-dashed and solid lines shown in Figure \ref{fig:SF_dist}, where the values of 
$S_{gr}$ are plotted. Here, the value of $\langle S_k\rangle (z)$ is computed by interpolating the mean $\langle S_{gr}\rangle$ for each redshift bin of 0.1. We see that our samples show almost no redshift dependencies. The value of $S_{gr, corr}$  is used to quantifies the colour variability behavior. If $S_{gr, corr} \approx 0$ within a 1$\sigma$ error, we take it as having no true colour variability trend, i.e., the brightness variations are at essentially constant colour. Whereas, $S_{gr, corr} < 0$  corresponds to a  bluer when brighter (BWB) trend, while $S_{gr, corr} > 0$ indicates that an object shows a redder when brighter (RWB) trend.  The BWB trend is found for the great majority of sources in both subclasses: NLSy1 (2013/2095) and BLSy1 (2257/2380), with the full breakdowns given in Table~\ref{tab:colr_variability_data}. 

Additionally, the quasi-simultaneous (within 0.5 hr) ZTF $g$- and $r$-band light curves of our sample can be used to study the pattern of colour index variation across different timescales using the method proposed by \citet{2014ApJ...792...54S}, defined as
\begin{equation}
\Theta(\tau) = \arctan\left( \frac{m_r(t + \tau) - m_r(t)}{m_g(t + \tau) - m_g(t)} \right),
\end{equation}
where $m_g$ and $m_r$ are the $g$- and $r$-band magnitudes at two epochs separated by a time lag $\tau$. Typically, quasars brighten or fade simultaneously in both bands, resulting in $\Theta \in [0^\circ, 90^\circ]$. However, due to photometric noise, $\Theta$ may also fall in the ranges $[-45^\circ, 0^\circ]$ or $[90^\circ, 135^\circ]$. We restrict $\Theta$ to the interval $[-45^\circ, 135^\circ]$ by transforming values outside this range through subtraction of $180^\circ$. The ensemble average of $\Theta$ at a given timescale is then computed as
\begin{equation}
\bar{\Theta}(\tau) = \frac{1}{N} \sum_{i=1}^{N} \Theta_i(\tau) ,
\end{equation}
where $N$ is the number of valid epoch pairs with time lag $\tau$.

We note here that in magnitude–magnitude space, the observed magnitudes represent the sum of the variable AGN and the constant host flux, resulting in a nonlinear relation. In such cases, the observed “bluer when brighter” (BWB) trend in total flux, measured within a finite aperture, can arise from the superposition of a constant red host galaxy component (including non-varying emission lines) and a variable blue AGN continuum \citep[e.g., see][]{Sakata2010ApJ...711..461S, 2015A&A...581A..93R}.

To account for this, we converted the $g$- and $r$-band magnitudes to corresponding flux values using the zero-point calibration provided by \citet{1999AJ....118.1406L}\footnote{\url{https://www.sdss4.org/dr17/algorithms/magnitudes/}}. We then computed the colour variation angle in flux–flux space, defined by \citet{2015A&A...581A..93R} as:
\begin{equation}
\theta(\tau) = \arctan\left( \frac{f_r(t + \tau) - f_r(t)}{f_g(t + \tau) - f_g(t)} \right)
\label{theta_flux}
\end{equation}
where $f_g$ and $f_r$ are the $g$- and $r$-band fluxes, respectively.

\begin{figure}
    \centering
    \includegraphics[width=0.46\textwidth, height=0.32\textwidth]{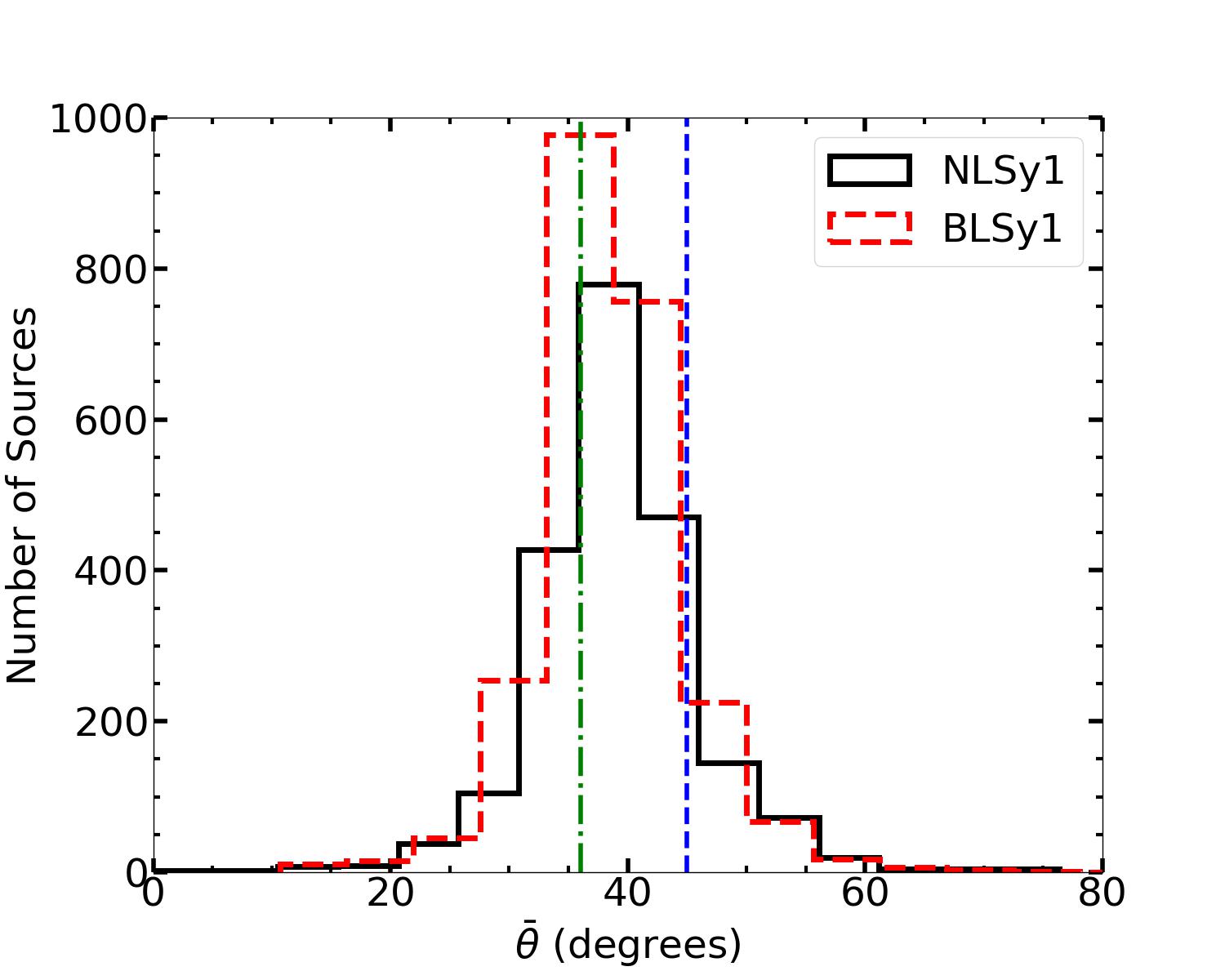}
    \caption{
        Distribution of $\bar{\theta}$ for NLSy1 (black bold line) and BLSy1 (red dashed line) sources using the flux--flux method (Eq.~\ref{theta_flux}). 
        The vertical dashed-dot green line at $36^\circ$ represents the median value for NLSy1 and BLSy1 populations, and the vertical dashed blue line at $45^\circ$ is shown only as a reference for $\Delta f_{r} = \Delta f_{g}$.
    }
    \label{fig:theta_distribution}
\end{figure}

\begin{figure*}
    \centering
    \includegraphics[width=0.46\textwidth, height=0.32\textwidth]{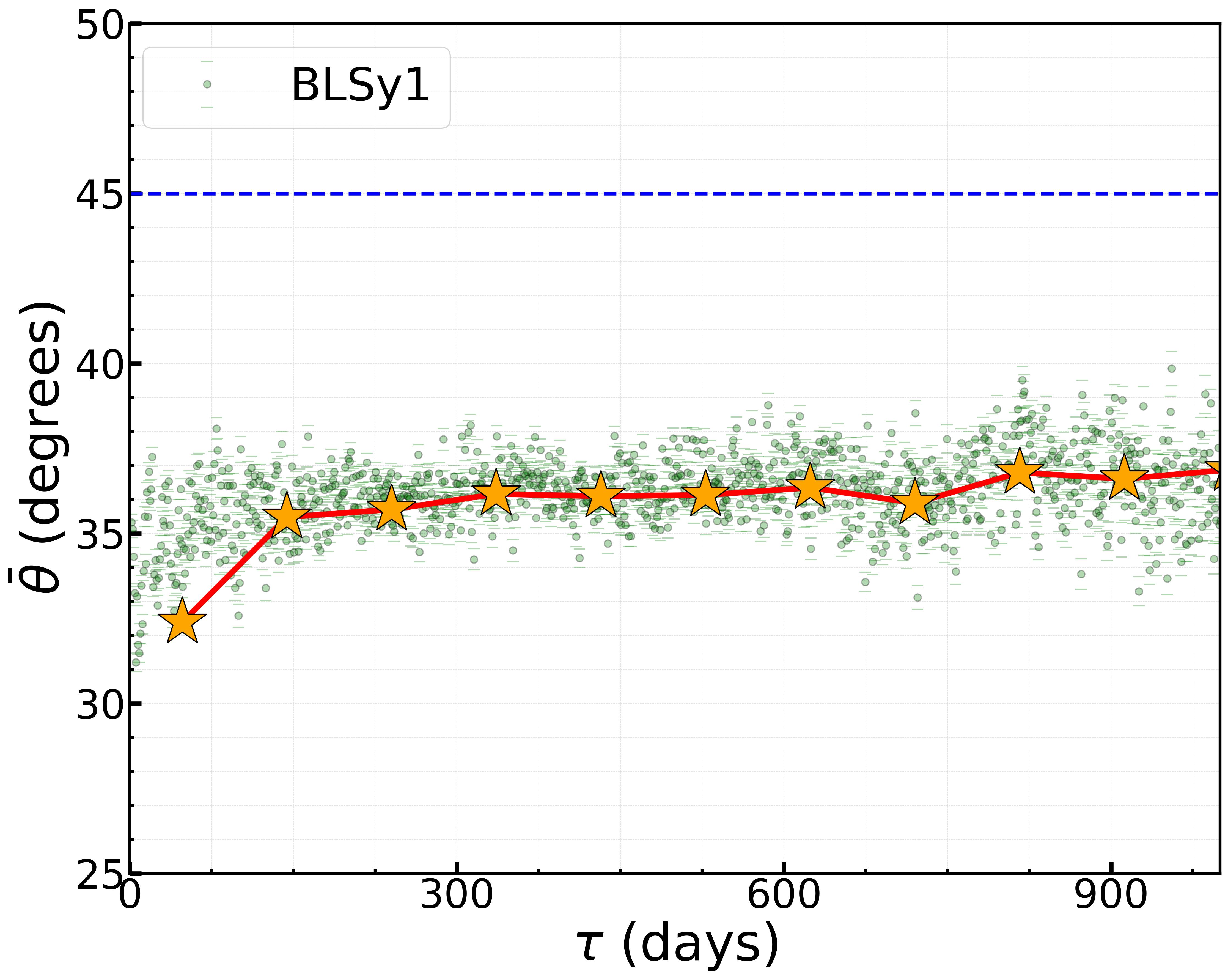}
    \includegraphics[width=0.46\textwidth, height=0.32\textwidth]{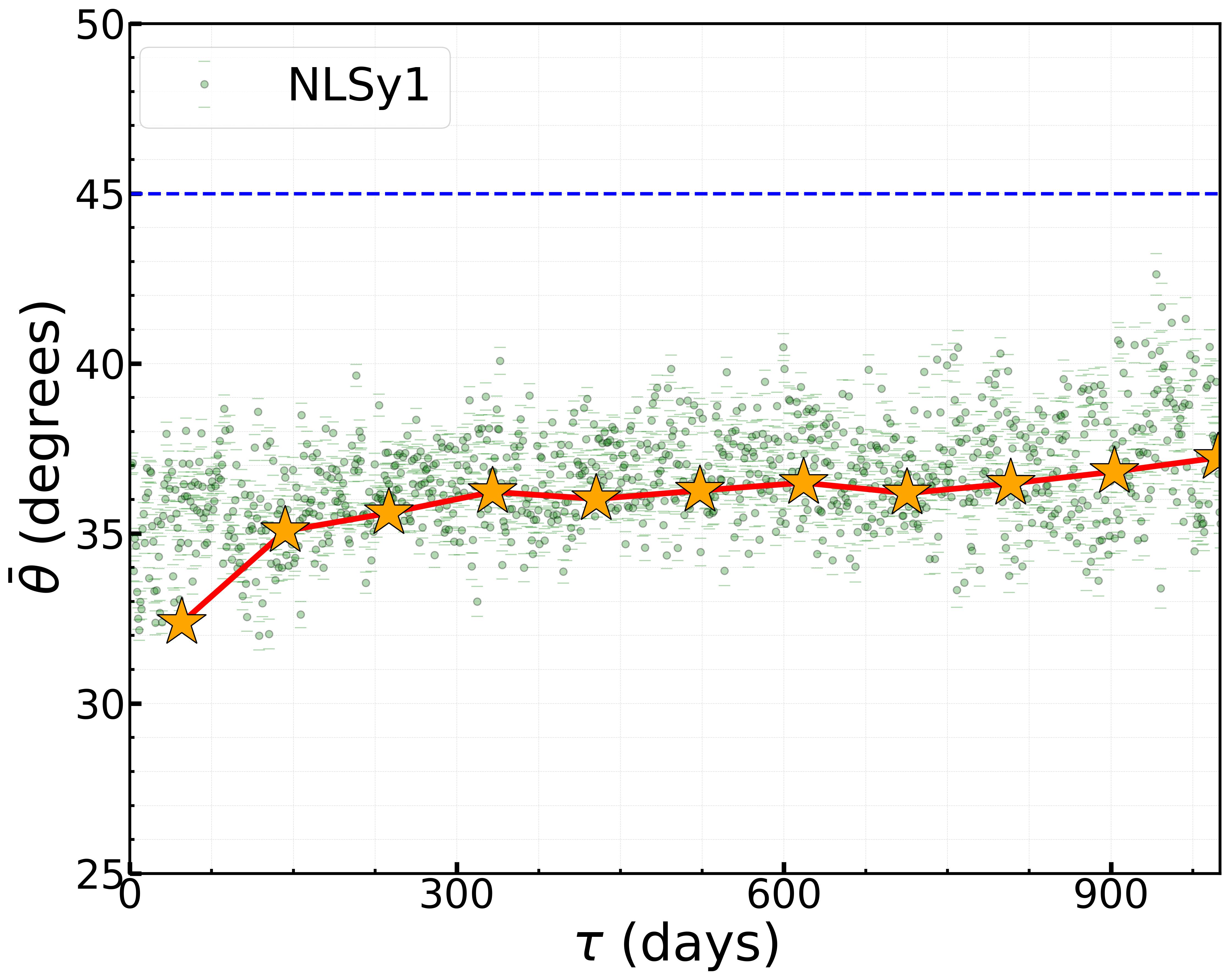}
    \caption{Distribution of $\bar{\theta}$ as a function of $\tau$ (rest-frame). The solid line represents the smoothed trend for each sample and the horizontal dashed line at $45^\circ$ is shown only as reference for $\Delta F_{r} = \Delta F_{g}$. 
    \textbf{\textit{Left:}} $\bar{\theta}$ variation for BLSy1 sources. 
    \textbf{\textit{Right:}} $\bar{\theta}$ variation for NLSy1 sources.}
    \label{fig:theta_variation}
\end{figure*}

 We estimated the colour variation angle $\theta$ for each source by averaging over all valid epoch pairs in the light curve. The distribution of average $\theta$ values for individual NLSy1 and BLSy1 sources is shown in Figure \ref{fig:theta_distribution}. 
 As seen in this figure, a majority of the sources have $\theta$ values clustered between $30^\circ$ and $45^\circ$, indicating a prevalent BWB trend in both subclasses: NLSy1 (1555 out of 2095) and BLSy1 (1870 out of 2380), with the full breakdown provided in Table~\ref{tab:colr_variability_data}. As shown in the table, using the flux–flux (magnitude–magnitude) method, the BWB trend is observed in 74.22\% (96.08\%) of the NLSy1 sample and 78.57\% (94.83\%) of the BLSy1 sample. The corresponding RWB trend is found in 7.68\% (0.28\%) of NLSy1s and 5.88\% (0.50\%) of BLSy1s using the same flux–flux (magnitude–magnitude) approaches. 
To further investigate the colour index variation as a function of timescale, we binned the epoch pairs by $\tau$ and computed the average $\theta$ across all sources for each bin. The resulting trends for both NLSy1 and BLSy1 populations are presented in Figure \ref{fig:theta_variation}, which show nearly constant colour index pattern observed across different timescales for both NLSy1 and BLSy1 galaxies. 

In the context of flux–flux analysis, we note that a constant average $\theta$ value, as seen in most of our sources, simply reflects the relative ratio of red to blue flux. For example, the classical Shakura–Sunyaev thin accretion disk model, which predicts $f_{\nu} \propto \nu^{1/3}$, corresponds to $\theta \approx 42^\circ$ for the SDSS $g$ and $r$ bands. In our sample, most sources exhibit $\theta$ values around $36^\circ$, which implies $f_{r}/f_{g} \approx 0.73$, indicating that the flux is predominantly bluer compared to the red flux.

\begin{figure*}
    \centering
    \includegraphics[width=0.49\textwidth,height=0.38\textwidth]{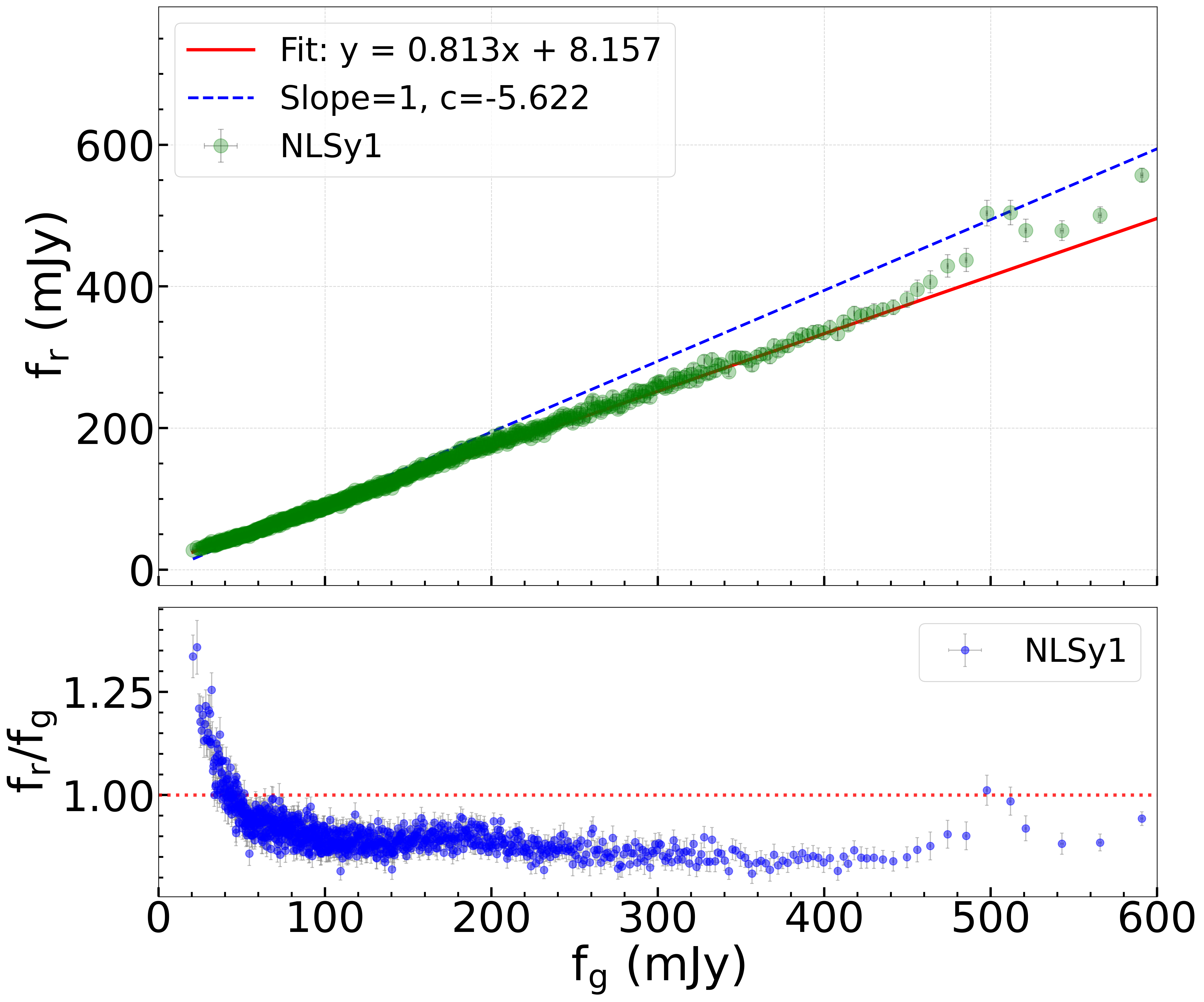}
    \includegraphics[width=0.49\textwidth,height=0.38\textwidth]{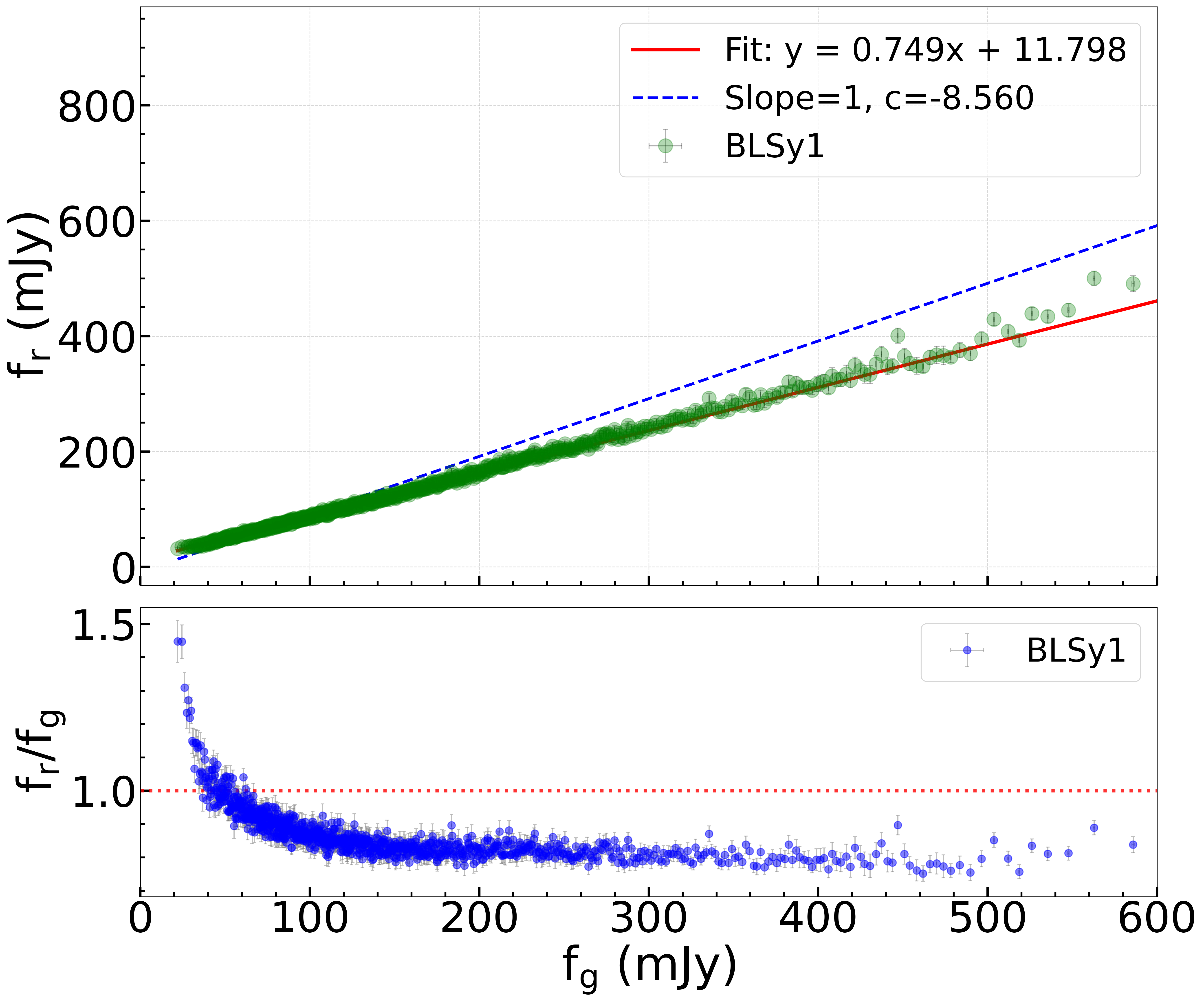}
    \caption{
    \textbf{\textit{Left panels (NLSy1):}} 
    Top: Weighted linear regression between $f_{r}$ and $f_{g}$ for NLSy1 galaxies. 
    The solid red line shows the best-fit relation $f_{r} = (0.813 \pm 0.002)\, f_{g} + (8.157 \pm 0.138)$, 
    while the dashed blue line represents a fixed slope of unity ($f_{r} = f_{g} + c$).  
    \textit{Bottom:} Flux ratio $f_{g}/f_{r}$ as a function of $f_{g}$, with propagated errors. 
    The red dotted line indicates the line where flux from both the $r$- and $g$-band are equal. 
    Error bars correspond to $1\sigma$ uncertainties. 
    \textbf{\textit{Right panels (BLSy1):}} 
    Top: Weighted linear regression between $f_{r}$ and $f_{g}$ for BLSy1 galaxies. 
    The solid red line shows the best-fit relation $f_{r} = (0.749 \pm 0.002)\, f_{g} + (11.798 \pm 0.146)$, 
    while the dashed blue line represents a fixed slope of unity ($f_{r} = f_{g} + c$).  
    Bottom: As for the left panel.}
    \label{fig:fr_fg_plot}
\end{figure*}

This result may partly be influenced by redshift effects, since for many sources the observed $g$ and $r$ passbands correspond to rest-frame ultraviolet wavelengths, where the bluer-when-brighter (BWB) behavior is theoretically expected to be stronger~\citep[see,][]{Netzer2025_10.1093/mnras/staf671}. To further explore this, we plot the ratio $f_r/f_g$ versus $f_g$ using quasi-simultaneous observations in the two bands (within half an hour), as shown in Figure \ref{fig:fr_fg_plot}. The figure reveals that this ratio decreases with $f_g$ for the ensemble average of both NLSy1 and BLSy1 galaxies. This may indicate the presence of a BWB trend. However, we caution that the trend could instead arise from an enhancement of the nuclear (variable) flux, which is predominantly bluer, relative to the nearly constant host galaxy contribution, which is predominantly redder. Detecting any genuine curvature in the flux–flux plots due to an intrinsic BWB trend of the nuclear flux will require careful subtraction of the host galaxy contribution and higher-quality data. 

\section{DISCUSSION} \label{DISCUSSION}
Variability of AGNs in brightness and colour have been important subjects of study for a long time. Several investigations have examined the brightness and colour variability behaviours of various subclasses of AGN (e.g., \citealp{2003MNRAS.345.1271V}; \citealp{2011PASJ...63..639I}; \citealp{2017ApJ...842...96R}; \citealp{2019MNRAS.483.2362R}; \citealp{2022MNRAS.510.1791N}).

Our study benefits from a statistically robust and well-defined pair of samples comprised of 2380 BLSy1 and 2095 NLSy1  
carefully matched in the $\mathrm{log(L_{Bol})}-z$ plane as discussed in Section \ref{Data_Preparation}. This matching ensures that any observed differences in variability properties arise primarily from intrinsic physical differences between the AGN subclasses rather than from selection biases. The use of multi-epoch light curves from ZTF DR22, which provides long-term coverage over 76 months with a median cadence of $\sim$3 days, offers an unparalleled opportunity to systematically investigate both brightness and colour variability. ZTF's extensive monitoring and high-quality photometry allow us to probe variability across different timescales, making it particularly well-suited for studying AGN variability in a statistically meaningful and temporally resolved manner.

\begin{figure}
	\includegraphics[width=0.46\textwidth,height = 0.32\textwidth]{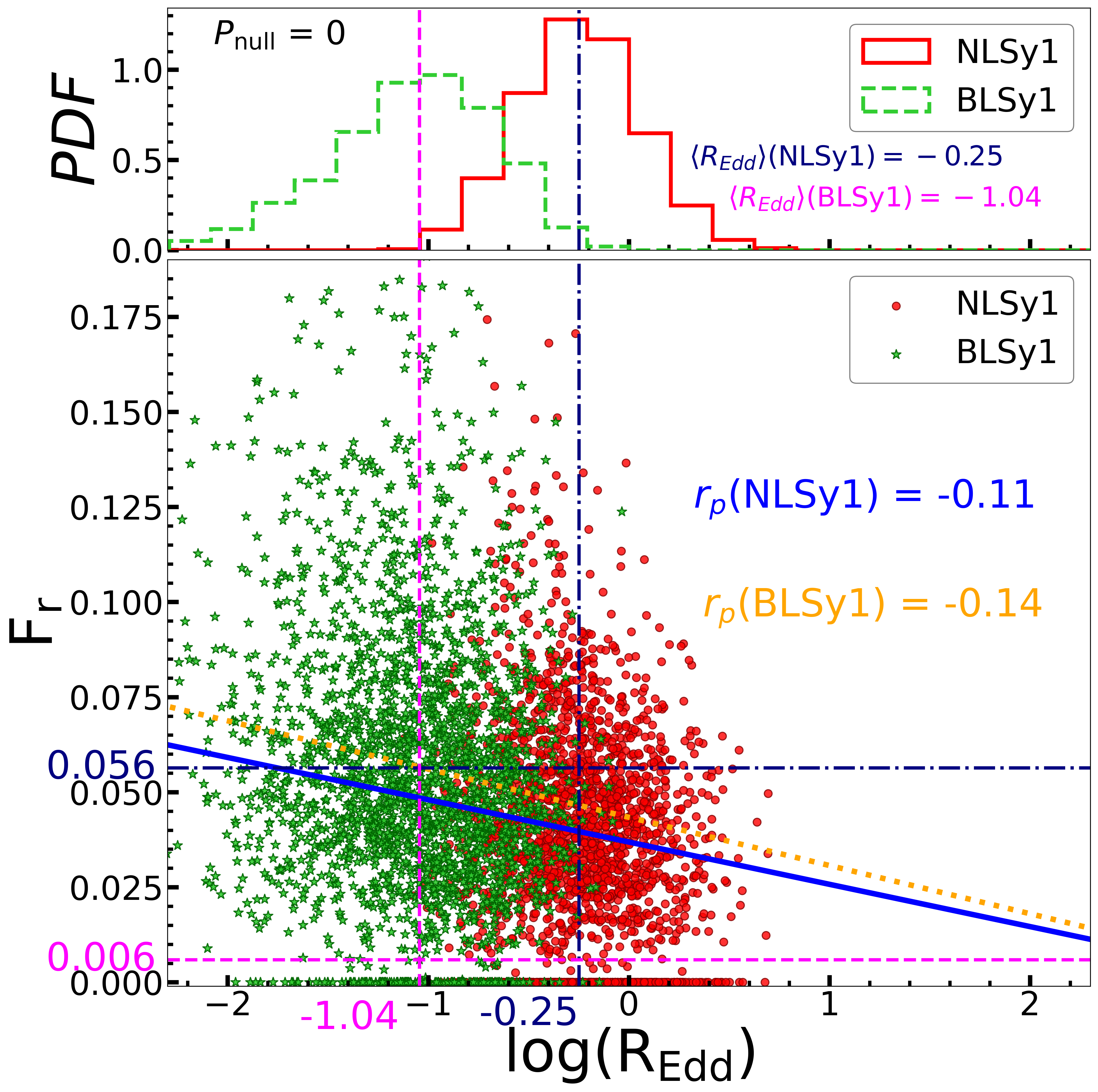}
    \caption{
The relationship between the Eddington ratio ($\log R_{\mathrm{Edd}}$) and the $\mathrm{r}$-band flux variability amplitude ($F_{r}$) for two distinct AGN populations, NLSy1 and BLSy1 galaxies. The \textit{upper panel} shows the distributions of $\log R_{\mathrm{Edd}}$ for both populations, with red solid-line histograms for NLSy1 and lime-green dashed-line histograms for BLSy1. Vertical lines denote median $\log R_{\mathrm{Edd}}$ values: dash-dotted navy for NLSy1 and dashed magenta for BLSy1. The KS2 test $p$-value, indicating the statistical difference between the distributions, is annotated in the upper panel. The lower panel presents the scatter plot of $F_{r}$ versus $\log R_{\mathrm{Edd}}$, where red circular markers with maroon edges denote NLSy1\_A sources, while green star markers with dark green edges represent BLSy1 sources. Linear regression fits for each population are overlaid: a solid blue line for NLSy1 and a dotted orange line for BLSy1. The Pearson correlation coefficients, denoted as $r_{p}$ for each AGN type, are also annotated. Vertical and horizontal dashed lines indicate the median value of $\log R_{\mathrm{Edd}}$ and corresponding $F_{r}$ values for both populations, with corresponding numerical values labeled.} 

    \label{fig:F_Redd_relationship}
\end{figure}

Comparing the  NLSy1 and BLSy1 samples, we find that BLSy1 sources exhibit significantly higher $\mathrm{r}$-band flux variability, 0.1144, and fractional flux variability $(2.615 \pm 0.161) \times 10^{-3}$, compared to 0.0826 and $(1.921 \pm 0.227) \times 10^{-3}$, respectively, for NLSy1 (Section~\ref{flux_frac}, Table \ref{tab:analysis_data}). Additionally, BLSy1 galaxies are found to have lower median Eddington ratios, $\log R_{\mathrm{Edd}}$
 of $-1.045$, compared to  $-0.249$ $\log R_{\mathrm{Edd}}$ of NLSy1. 
This overall anti-correlation between flux variability and $\log R_{\mathrm{Edd}}$ is illustrated in Figure~\ref{fig:F_Redd_relationship}.
We note that the adopted values of $M_{\mathrm{BH}}$ used to estimate $\log R_{\mathrm{Edd}}$ are subject to significant scatter. This is primarily because the $M_{\mathrm{BH}}$ estimates by \citet{2024MNRAS.527.7055P}, although based on the updated $R$–$L$ relation from \citet{2019ApJ...886...42D}, rely on the single-epoch method, in which the black hole mass is computed using the formula $M_{\mathrm{BH}} = f R V^2 / G$. Here, $R$ is the radius of the broad-line region (BLR), i.e., the distance between the BLR clouds emitting a particular broad emission line and the central continuum source, and $V$ is the velocity dispersion of the line-emitting gas, inferred from the width of the line, typically using either the full width at half maximum (FWHM) or line dispersion, $G$ is the gravitational constant, and the scaling factor $f$ accounts for the geometry and kinematics of the BLR, and is taken as $f = 0.75$ assuming an isotropic or spherical distribution of BLR clouds (e.g., \citealp{2017ApJS..229...39R}). However, this assumption can introduce considerable uncertainties in the mass estimates when compared to the more reliable reverberation mapping method, which directly measures the time lag between the continuum and line emission to estimate $R$, but can only be carried out for a small fraction of AGN (e.g., \citealp{BontaPeterson}). Accounting for both measurement and systematic uncertainties, the total uncertainty in these single-epoch estimates can be as large as $\sim 0.4$ dex (see also \citealp{2013BASI...41...61S}), and may introduce additional uncertainty in the anti-correlation we found between flux variability and $\log R_{\mathrm{Edd}}$ in Figure~\ref{fig:F_Redd_relationship}.

Moreover,  the offset in $\log R_{\mathrm{Edd}}$ between NLSy1 and BLSy1 in Figure \ref{fig:F_Redd_relationship} primarily results from our sample matching in the $z$–$L_{\mathrm{Bol}}$ plane and the use of line-width-based black hole mass estimates. Since NLSy1s are defined by narrower H$\beta$ lines, they naturally yield higher Eddington ratios under this method. Our goal with Figure \ref{fig:F_Redd_relationship} is merely to explore variability trends within matched subsets without claiming they can be generalised to the full NLSy1 population 
\citep[e.g.,][]{2017ApJ...842...96R}.

Both of our classes, NLSy1 and BLSy1, exhibit significant amplitude of temporal variability, with magnitudes of 0.77 and 0.83 
respectively (Section~\ref{ANALYSIS_AND_RESULTS}). The differing variability amplitudes between these classes suggest that distinct physical mechanisms govern their variability. 
The higher variability in BLSy1s compared to NLSy1s could be attributed to increased thermal emission from their accretion discs.
This result is consistent with the previous finding by \citet{2017ApJ...842...96R}, where they have used DRW modelling to analyze the optical variability of 5510 NLSy1 and a control sample of 5510 BLSy1 using CRTS light curves \citep[see also][]{2004ApJ...609...69K, 2010ApJ...716L..31A, 2013AJ....145...90A}. One possible explanation is the presence of slim accretion discs \citep{1988ApJ...332..646A} in NLSy1 galaxies compared to the standard Shakara-Sunyaev, geometrically thin, optically thick, accretion disc in BLSy1 galaxies \citep{2013AJ....145...90A}. The slim disc scenario is compatible with the fact that the mean Eddington ratio of NLSy1s is significantly higher than that of BLSy1s. Then, the lower variability of NLSy1 is consistent with the previously found anti-correlation of AGN variability with the Eddington ratio \citep{2008MNRAS.383.1232W, 2010ApJ...716L..31A, 2010ApJ...721.1014M}. Hence, the reprocessing of X-ray emissions into optical/UV bands may not be as significant in NLSy1s as in BLSy1s \citep{2013AJ....145...90A}. This could imply that the intrinsic variability in NLSy1s is more directly related to their accretion processes, rather than being heavily influenced by reprocessed emissions, leading to a more pronounced variability in the X-ray regime but less variability in the optical regime.
 
As discussed in Section~\ref{SF_sec}, the structure function (SF) is particularly effective for comparing the variability characteristics of AGN classes over a range of timescales, especially when dealing with irregularly sampled light curves. Our SF analysis shows that the median value of the SF amplitude, over timescales ranging from 20 to 1000 days, for BLSy1 galaxies is $1.44 \pm 0.06$ times higher than for NLSy1 (left panel of Figure~\ref{fig:SF_dist_BLSy1}). 
The enhanced variability across all timescales observed in BLSy1 galaxies relative to NLSy1s may be consistent with a more direct line of sight to their compact, rapidly varying nuclear regions. For NLSy1s, reduced variability amplitudes could arise from a variety of factors, such as possible additional obscuration along the line of sight or geometrical dilution effects, although these remain speculative and require further investigation.

The slope of the structure function describes how rapidly variability increases with time lag. The best-fit SF slopes in $\log \mathrm{SF}$--$\log \Delta T$ space for our samples are: $0.24 \pm 0.02$ for BLSy1 and $0.25 \pm 0.02$ for NLSy1, which are consistent within $1\sigma$ (right panel of Figure~\ref{fig:SF_dist_BLSy1}). These values are broadly consistent with the QSO SF slope of $\sim 0.20 \pm 0.01$ reported by \citet{2002MNRAS.329...76H}. The slight discrepancy likely arises from the fact that our control samples were matched in the $\log(L_{\mathrm{Bol}})$--$z$ plane.

Using photometric data in $\mathrm{g}$- and $\mathrm{r}$-bands from the ZTF survey, we find that large majorities of the sources in all our samples  exhibit a BWB trend, as summarized in Table \ref{tab:colr_variability_data}. As noted in Section~\ref{colour_var_sec}, the observed flux in our NLSy1 and BLSy1 samples includes contributions from both the variable AGN component and the non-variable host galaxy within the finite aperture. This combination leads to a non-linear relationship in logarithmic magnitude–magnitude space; hence, we use the flux–flux space to determine the spectral index and its variation. Due to the large size of our sample, we could not isolate the host galaxy fluxes. Consequently, the observed BWB trend may result from using total fluxes (e.g, see Eq.~\ref{theta_flux}), where a typically blue AGN brightens while the redder host galaxy remains constant (see also \citealt{Sakata2010ApJ...711..461S, 2015A&A...581A..93R}). Alternatively, the nucleus itself may intrinsically become bluer during bright phases, particularly in the UV, though such effects are generally weak and difficult to detect in the optical \citep[e.g.,][]{Pereyra2006ApJ...642...87P, Sakata2010ApJ...711..461S}. Our analysis of the colour index variation across different timescales reveals a nearly constant trend for both NLSy1 and BLSy1 galaxies (see, Figure \ref{fig:theta_variation}), with the variable component being significantly bluer than predicted by a standard Shakura–Sunyaev disk model (see, Section~\ref{colour_var_sec}). While such variability could plausibly originate from a hot thermal region of the accretion disk \citep{2015A&A...581A..93R}, additional effects may also contribute. In particular, the relatively high redshift of many sources shifts the observed g and r passbands into the rest-frame ultraviolet, where a stronger bluer-when-brighter behavior is expected \citep[e.g., see][]{Netzer2025_10.1093/mnras/staf671}, suggesting that such an effect may also be at play.
A similar colour variability investigation of NLSy1s in the mid-infrared was performed by \citet{2019MNRAS.483.2362R}; however, such infrared studies may be dominated by dust emission, while the optical emission comes predominantly from the accretion disc.

Several studies have investigated the colour variability behaviour of blazars (e.g., \citealt{2003ApJ...590..123V, 2011PASJ...63..639I, 2022MNRAS.510.1791N}). 
When examining BL Lacs and FSRQs separately, these studies reported that BL Lacs predominantly follow a BWB trend, while the majority of FSRQs exhibit the RWB trend. A possible explanation for this contrasting behaviour was recently discussed by \citet{2022MNRAS.510.1791N}, who noted that in BL Lacs, non-thermal synchrotron emission from the jet dominates the optical continuum, resulting in the observed BWB trend. In contrast, the RWB trend seen in FSRQs indicates a significant contribution from quasi-thermal emission from the accretion disc, with a relatively less important contribution from the jet.

\section{Conclusions} \label{Conclusions}

To try to better understand the physical mechanisms governing the variability around the central engine, we studied the optical brightness and colour brightness variability of NLSy1 sources and their corresponding $\mathrm{log(L_{Bol})}$-$z$ matched BLSy1 sources. 
In this work we have used $\mathrm{r}$- and $\mathrm{g}$-band lightcurve data from the ZTF DR22 survey which spans over 6 years ($\sim 76$ months). We have used a sample of 2,095 NLSy1 sources matched with 2,380 BLSy1 sources, ensuring close agreement in both redshift ($z$) and bolometric luminosity ($\log L_{\text{Bol}}$), with a matching tolerance of 0.001 in each parameter. Each source involved in this study has at least 100 light curve data points; quasi-simultaneous light curves with half-hour epoch differences were employed to investigate colour variability.  Therefore, the present samples, along with the well constrained and rich dataset, are ideal for a comparative study of the optical brightness and colour variability properties of the NLSy1 and BLSy1 samples. 
We have analyzed the comparative flux, fractional flux, amplitude and colour variability of these sources along with their structure functions. Our main findings of this work are as follow:

(1) The $\mathrm{r}$-band median flux variability  is 0.0826 for NLSy1 and 0.1144 for BLSy1 indicating that BLSy1s exhibit greater flux variability than NLSy1s. Similarly, the fractional flux variability is found to be 0.002 and  0.003 for NLSy1 and BLSy1, respectively.  BLSy1 sources show higher fractional flux variability ($F_{var}$) than NLSy1 sources  matched to them. In this case, KS2 tests confirm that NLSy1 sources are significantly different from BLSy1.

(2) Based on the computed $\Delta F_{\text{var,gr}} = F_{\text{var,g}} - F_{\text{var,r}}$, we concluded that the fractional variability ($F_{\text{var}}$) is effectively frequency-independent across the $\mathrm{r}$- and $\mathrm{g}$-bands in our sample.

(3) Both NLSy1 and BLSy1 exhibit different amplitudes of temporal variability over the  $\sim 6$ years timescale, with magnitudes of 0.77 and 0.83, 
respectively. This clearly indicates that BLSy1 sources show greater amplitude of temporal variability than NLSy1.  
KS2 tests confirm that NLSy1 sources are significantly different from BLSy1.

(4) Our analysis shows that the $\langle \mathrm{SF} \rangle$ value for BLSy1 sources is $1.44 \pm 0.06$ times higher than that of NLSy1s.
Our best fit SF slope parameter in $\mathrm{log\,SF} - \mathrm{log}\,\Delta T$ space is found to be 0.24 for NLSy1 and 0.25 for BLSy1.

(5) Our colour variability analysis reveals that the great majority of sources in all classes follow the BWB trend: NLSy1 (1555/2095) and BLSy1 (1870/2380). We observe a nearly constant colour index variation across timescales for both NLSy1 and BLSy1 galaxies.

Collectively, these findings offer important clues about the  accretion processes operating in NLSy1 and BLSy1 sources. 
However, further simultaneous multiwavelength studies, incorporating both higher-cadence data and broader spectral coverage, will be necessary to more fully unravel the underlying physical mechanisms driving these trends.

\section*{Acknowledgements}
We thank the referee Prof. Hartmut Winkler for constructive comments and suggestions, which helped to improve the clarity and quality of this manuscript. This research of M.S. is supported by the University Grants Commission (UGC), Government of India, under the UGC-JRF/SRF scheme  (Ref.\ No.: 221610066226). H.C., M.S. and R.K. express their gratitude to the Inter-University Centre for Astronomy and Astrophysics (IUCAA) for their hospitality and the provision of High-Performance Computing (HPC) facilities under the IUCAA Associate Programme. 

This work is based on observations obtained with the Samuel Oschin 48-inch Telescope and the 60-inch Telescope at the Palomar Observatory as part of the Zwicky Transient Facility (ZTF) project. ZTF is supported by the National Science Foundation under Grants No. AST-1440341 and AST-2034437 and a collaboration including current partners Caltech, IPAC, the Oskar Klein Center at Stockholm University, the University of Maryland, University of California Berkeley, the University of Wisconsin at Milwaukee, University of Warwick, Ruhr University, Cornell University, Northwestern University, and Drexel University. Operations are conducted by COO, IPAC, and the University of Washington.

\section*{Data Availability}
The NLSy1 and BLSy1 catalogs used in this study were extracted from \citet{2024MNRAS.527.7055P} 
 and the light curve data are taken from the publicly available ZTF DR22 \citep{2019PASP..131a8002B}.



\bibstyle{mnras}
\bibliography{biblography.bib} 








\bsp	
\label{lastpage}
\end{document}